\documentclass[conference]{IEEEtran}

\usepackage{color}
\usepackage[english]{babel}
\usepackage{booktabs}
\usepackage{paralist}
\usepackage{graphicx}
\usepackage{multirow}
\usepackage{url}
\usepackage{float}
\usepackage{cite}
\usepackage{enumerate}
\usepackage{enumitem}
\usepackage{fixltx2e}
\usepackage{algorithm}
\usepackage[noend]{algpseudocode}
\usepackage{relsize}
\usepackage{subfig}
\usepackage{texdef2015}
\usepackage{setspace}
\usepackage{balance}
\newcommand{\age}{\Delta}

\usepackage{amstext} 
\usepackage{array}   
\newcolumntype{L}{>{$}l<{$}} 

\def\BibTeX{{\rm B\kern-.05em{\sc i\kern-.025em b}\kern-.08em
    T\kern-.1667em\lower.7ex\hbox{E}\kern-.125emX}}

\begin{document}


\title{ACP: An End-to-End Transport Protocol for Delivering Fresh Updates in the Internet-of-Things}

\author{\IEEEauthorblockN{Tanya Shreedhar}
\IEEEauthorblockA{\textit{Wireless Systems Lab, IIIT-Delhi} \\
tanyas@iiitd.ac.in}
\and
\IEEEauthorblockN{Sanjit K. Kaul}
\IEEEauthorblockA{\textit{Wireless Systems Lab, IIIT-Delhi} \\
skkaul@iiitd.ac.in}
\and
\IEEEauthorblockN{Roy D. Yates}
\IEEEauthorblockA{\textit{WINLAB, Rutgers University} \\
ryates@winlab.rutgers.edu}
}

%
\maketitle
%
\IEEEpubidadjcol



\begin{abstract}

The next generation of networks must support billions of connected devices in the Internet-of-Things (IoT). To support IoT applications, sources sense and send their measurement updates over the Internet to a monitor (control station) for real-time monitoring and actuation. Ideally, these updates would be delivered at a high rate, only constrained by the sensing rate supported by the sources. However, given network constraints, such a rate may lead to delays in delivery of updates at the monitor that make the freshest update at the monitor unacceptably old for the application.

We propose a novel transport layer protocol, namely the Age Control Protocol (ACP), that enables timely delivery of such updates to monitors, in a network-transparent manner. ACP allows the source to adapt its rate of updates to dynamic network conditions such that the average age of the sensed information at the monitor is minimized. We detail the protocol and the proposed control algorithm. We demonstrate its efficacy using extensive simulations and real-world experiments. To exemplify, ACP achieves about 100 msec of reduction in age in comparison to a baseline that has sources send one update every round-trip-time (RTT). This is for multiple sources that send their updates over the Internet to monitors on another continent over an end-to-end link with round-trip-times of 185 msec.

\end{abstract}  


\section{Introduction}
\label{sec:introduction}
The availability of inexpensive embedded devices with the ability to sense and communicate has led to the proliferation of a relatively new class of real-time monitoring systems for applications such as health care, smart homes, transportation, and natural environment monitoring. Devices repeatedly sense various physical attributes of a region of interest, for example, traffic flow at an intersection. This results in a device (the \emph{source}) generating a sequence of packets (\emph{updates}) containing measurements of the attributes. A more recently generated update contains a more current measurement. The updates are communicated over the Internet to a \emph{monitor} that processes them and decides on any actuation that may be required. 

For such applications, it is desirable that freshly sensed information is available at monitors. However, as we will see, simply generating and sending updates at a high rate over the Internet is detrimental to this goal. In fact, freshness at a monitor is optimized by the source smartly choosing an update rate, as a function of the end-to-end network conditions. Freshness at the monitor suffers when a too small or a too large rate of updates is chosen by the source. In this work, we propose the Age Control Protocol (ACP), which in a network-transparent manner regulates the rate at which updates from a  source are sent over its end-to-end connection to the monitor. This rate is such that the average age, where the age of an update is the time elapsed since its generation by the source, of sensed information at the monitor is kept to a minimum, given the network conditions. Based on feedback from the monitor, ACP adapts its suggested rate to the perceived congestion in the Internet. Consequently, ACP also limits congestion that would otherwise be introduced by sources sending to their monitors at unnecessarily fast update rates. 

The requirement of freshness is not akin to requirements of other pervasive real-time applications like voice and video. For these applications, the rate at which packets are sent is determined by the codec being used. Often, such applications adapt to network conditions by choosing an appropriate code rate. These applications, while resilient to packet drops to a certain degree, require end-to-end packet delays to lie within known limits and would like small end-to-end jitter. Monitoring applications may achieve a low update packet delay by simply choosing a low rate at which the source sends updates. This, however, may be detrimental to freshness, as a low rate of updates can lead to a large \emph{age} of sensed information at the monitor, simply because updates from the source are infrequent. More so than voice/video, monitoring applications are exceptionally loss resilient and they don't benefit from the source retransmitting lost updates. Instead, the source should continue sending new updates at its configured rate.
\begin{figure}[!t]             
\begin{center}
\includegraphics[width=0.4\textwidth]{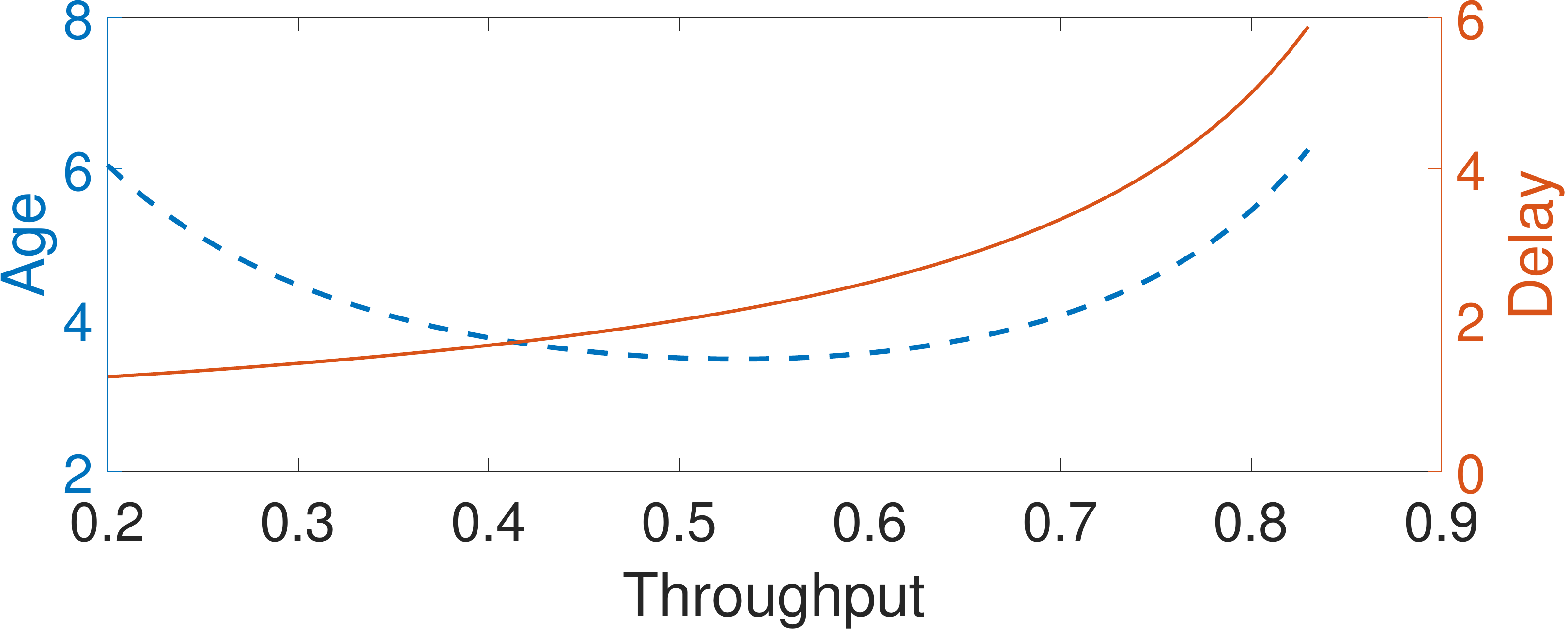}
\caption{Interplay of the networking metrics of delay (solid line), throughput (normalized by service rate) and age. Shown for a M/M/1 queue~\cite{shortle2018fundamentals} with service rate of $1$. The age curve was generated using the analysis for a M/M/1 queue in~\cite{KaulYatesGruteser-Infocom2012}.}
\label{fig:networkingMetrics}
\end{center}
\vspace{-0.2in}
\end{figure}

At the other end of the spectrum are applications like that of file transfer that require reliable transport and high throughputs but are delay tolerant. Such applications use the transmission control protocol (TCP) for end-to-end delivery of application packets. As we show in Section~\ref{sec:TCP}, the congestion control algorithm of TCP, which optimizes the use of the network pipe for throughput, and TCP's emphasis on guaranteed and ordered delivery is detrimental to keeping age low.

Figure~\ref{fig:networkingMetrics} broadly captures the behavior of the metrics of delay and age as a function of throughput. Under light and moderate loads when packet dropping is negligible, throughput (average network utilization) increases linearly in the rate of updates. This leads to an increase in the average packet delay. Large packet delays coincide with large average age. Large age is also seen for small throughputs (and corresponding small rate of updates). \emph{At a low update rate, the monitor receives updates infrequently, and this increases the average age (staleness) of its most fresh update}. Finally, observe that there exists a sending rate (and corresponding throughput) at which age is minimized.

Many works, see Section~\ref{sec:related}, have analyzed age as a Quality-of-Service metric for monitoring applications. Often such works have employed queue theoretic abstractions of networks. More recently in~\cite{sert2018DeepQ} the authors proposed a deep Q-learning based approach to optimize age over a given but unknown network topology. We believe our work is the first to investigate age control at the transport layer of the networking stack, that is over an end-to-end connection in an IP network and in a manner that is transparent to the application. A preliminary version of this work appeared as a poster~\cite{shreedhar2018acp}. Our specific contributions are listed next.\\

\noindent \textbf{(a)} We propose the Age Control Protocol, a novel transport layer protocol for real-time monitoring applications that wish to deliver fresh updates over IP networks. ACP regulates the rate at which a status updating source sends its updates to a monitor over its end-to-end connection in a manner that is application independent and makes the network transparent to the source. We argue that such a protocol, unlike other transport protocols like TCP and RTP, must have just the right number of update packets in transit at any given time. 

\noindent \textbf{(b)} We demonstrate that TCP, which is the most commonly used transport protocol in the Internet is not suitable for the transport of update packets. 

\noindent \textbf{(c)} We provide an extensive evaluation of the protocol using network simulations and real world experiments in which one or more sources sends packets to monitors. While our simulations are constrained to six hop paths between the sources and the monitors, in our real experiments we have sources send their updates over an inter-continental end-to-end IP connection. Over such a connection with a median round-trip-time of about $185$ msec, ACP achieves a significant reduction in median age of about $100$ msec ($\approx 33\%$ improvement) over age achieved by a protocol that sends one update every RTT.

The rest of the paper is organized as follows. In the next section, we describe related works. In Section~\ref{sec:TCP} we demonstrate why the mechanisms of TCP are detrimental to minimizing age. In Section~\ref{sec:ACPProtocol}, we detail the Age Control Protocol, how it interfaces with a source and a monitor, and the protocol's timeline. In Section~\ref{sec:problem} we define the age control problem. In Section~\ref{sec:acpIntuit}, we use simple queueing models to intuit a good age control protocol and discuss a few challenges. Section~\ref{sec:ACPProtocol} details the control algorithm that is a part of ACP. This is followed by details on the evaluation methodology in Section~\ref{sec:evaluation}. We discuss simulation results in Section~\ref{sec:simulationResults} and results from real-world experiments in Section~\ref{sec:realWorldResults}. We conclude in Section~\ref{sec:conclusions}.

\section{Related Work}
\label{sec:related}
The need for timely updates arises in many fields, including, for example,  vehicular updating \cite{kaul_minimizing_2011}, real time databases \cite{xiong_deriving_1999}, data warehousing \cite{karakasidis_etl_2005}, and web caching \cite{yu_scalable_1999,Cho-GM-TODS2003effective}.   

For sources sending updates to monitors,
there has been growing interest in the age of information (AoI) metric that was first analyzed for elementary queues in \cite{KaulYatesGruteser-Infocom2012}. 
To evaluate AoI for a single source sending updates through a network cloud \cite{KamKompellaEphremides2013ISIT} or through an M/M/k server \cite{KamKompellaEphremides2014ISIT,KamKNE2016IT}, out-of-order packet delivery was the key analytical challenge. 
Packet deadlines are  found to improve AoI in \cite{KamKompellaNWE-ISIT2016}.  AoI in the presence of errors is evaluated in \cite{ChenHuang-ISIT2016}. Distributional properties of the age process have also been analyzed for the D/G/1 queue under first-come-first-served (FCFS) \cite{Champati-AG-AOI2018}, as well as   single server FCFS and LCFS queues  \cite{Inoue-MTT-arxiv2017}. There have also been studies of energy-constrained updating \cite{Elif2015ITA,Yates2015ISIT,UpdateorWait-IT2017,Nath-WY-spawc2017,FaraziKleinBrown-AOI2018,Arafa-YU-ICC2018}.

There has also been substantial efforts to evaluate and optimize age for multiple sources sharing a communication link
\cite{Kadota-UBSM-Allerton2016,KaulYates-ISIT2017,Sang-LJ-globecom2017, NajmTelatar-AOI2018,JiangKrishnmachariZZN-ISIT2018}.  In particular, near-optimal scheduling based on the Whittle index has been explored in \cite{Kadota-UBSM-Allerton2016,Jiang-KZN-itc2018,Hsu-ISIT2018}. 
When multiple sources employ wireless networks subject to interference constraints, AoI has been analyzed under a variety of link scheduling methods \cite{LuJiLi-Mobicom2018,Talak-KKM-arxiv2018}. AoI analysis for multihop networks has also received attention \cite{TalakKaramanModiano-Allerton2017}. Notably, optimality properties of a Last Generated First  Served (LGFS) service when updates arrive out of order are found in \cite{Bedewy-SS-arxiv2017}. 

While the early work \cite{kaul_minimizing_2011} explored  practical issues such as contention window sizes, the subsequent AoI literature has primarily been focused on analytically tractable simple models. 
Moreover,  a model for the system is typically assumed to be known. 
In this work, our objective has been to develop end-to-end updating schemes that perform reasonably well without assuming a particular network configuration or model. This approach attempts to learn (and adapt to time variations in) the condition of the network links from source to monitor.  This is similar in spirit to  hybrid ARQ based updating schemes \cite{Ceran-GG-WCNC2018,NajmYatesSoljanin-ISIT2017} that learn the wireless channel. The chief difference is that hybrid ARQ occurs on the short timescale of a single update delivery while ACP learns what the network supports over many delivered updates.

\begin{figure*}[!th]
	\begin{center}
		\subfloat[]{\includegraphics[width=.33\textwidth]{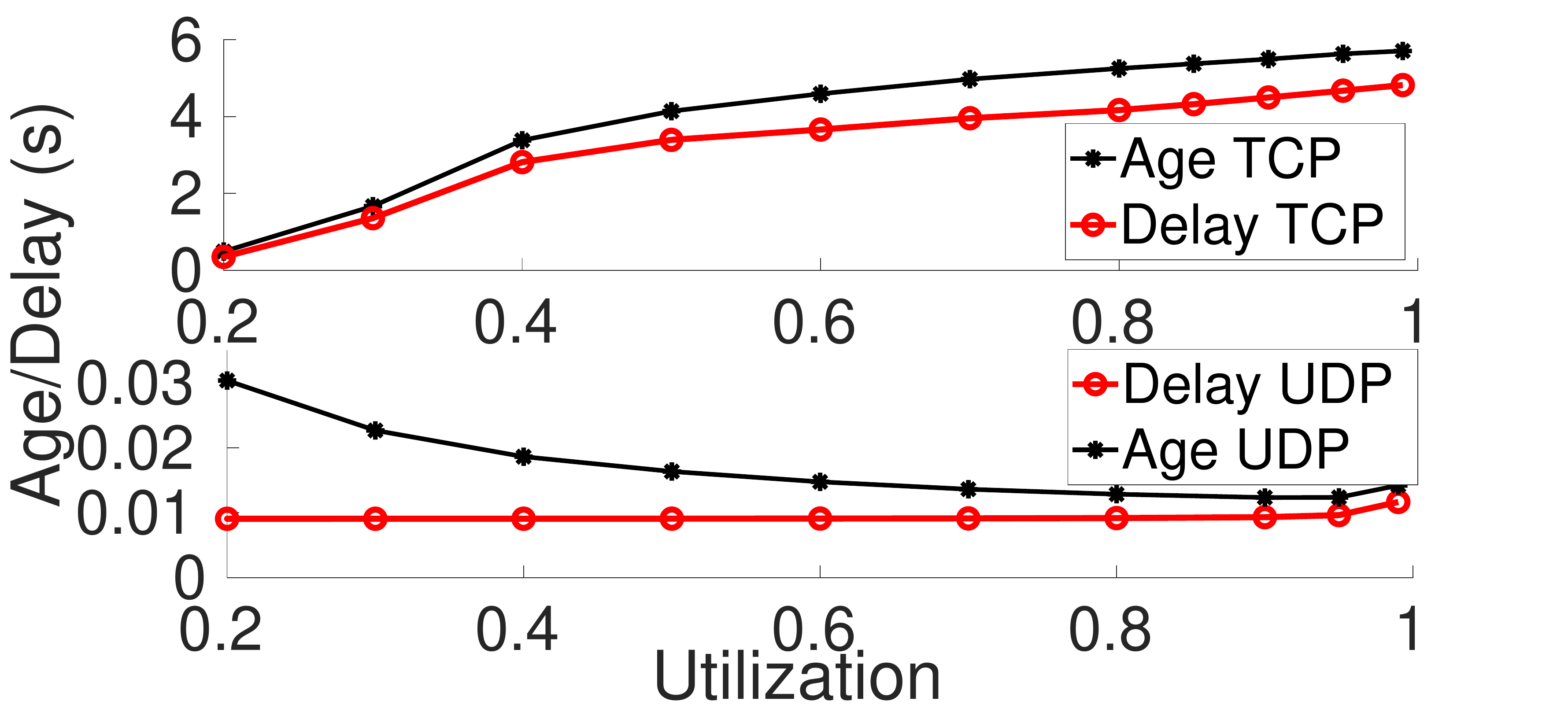}
			\label{fig:UDPvsTCPPacketError}}
		\subfloat[]{\includegraphics[width=.33\textwidth]{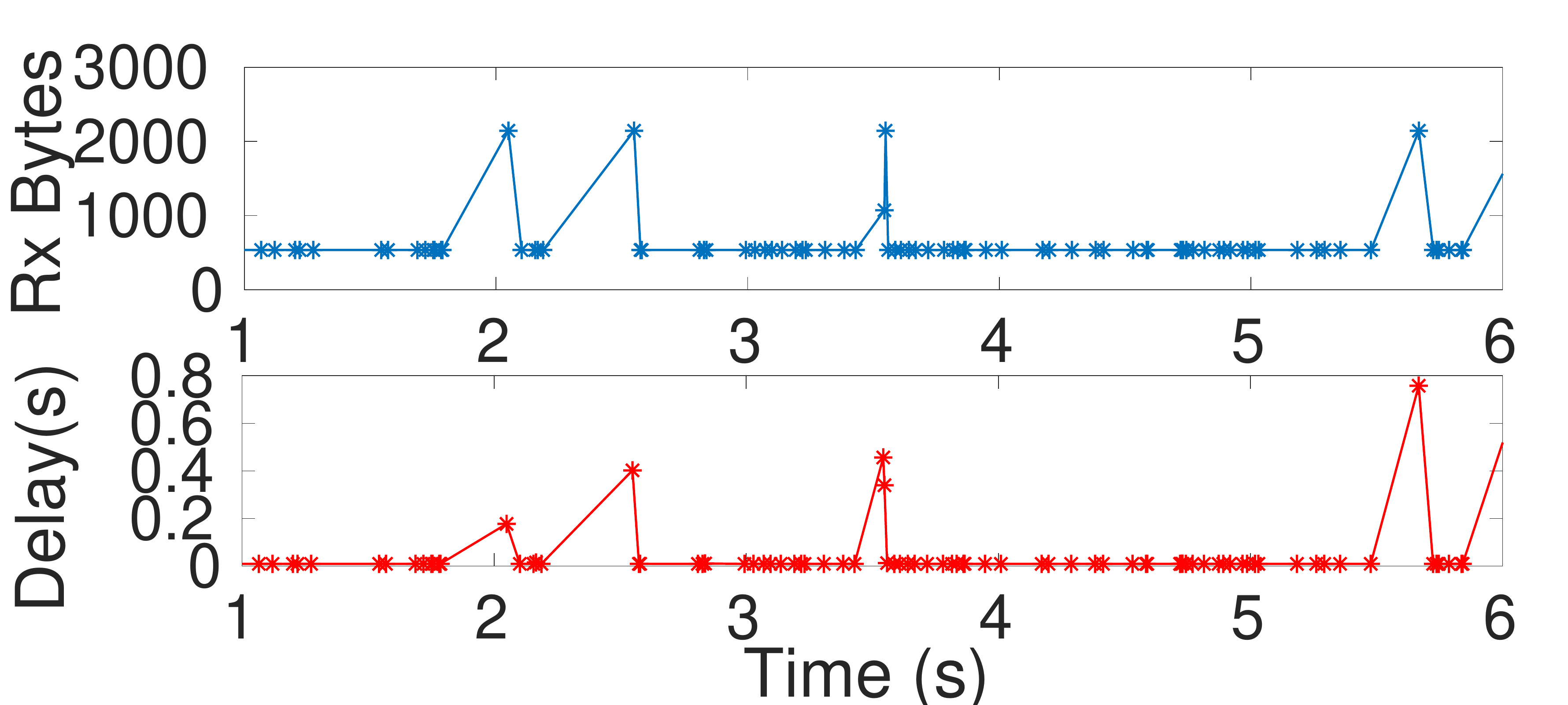}
			\label{fig:UDPvsTCPPacketErrorRXBuffer}}
		\subfloat[]{\includegraphics[width=.33\textwidth]{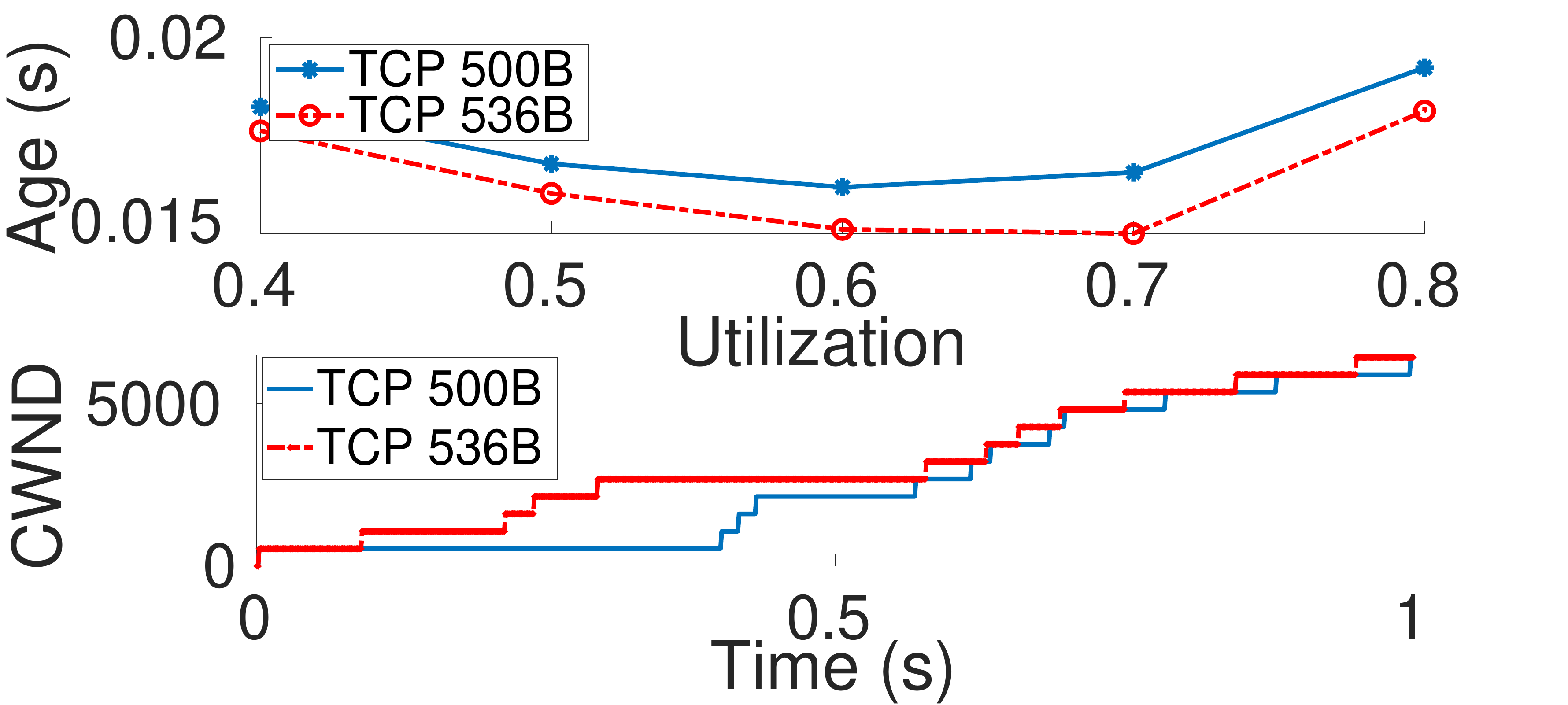}
			\label{fig:TCPSmallPacketSizes}}
		\label{fig:TCPEval}
		\caption{(a) Impact of packet errors on age and packet delay when using TCP and UDP (b) A time snapshot of delays suffered by update packets transmitted using TCP and the corresponding received bytes by the monitor. Packet error rate was set to $0.1$ (c) Age as a function of packet size and how packet size impacts increase of the TCP congestion window.}
	\end{center}
\vspace{-0.2in}
\end{figure*} 
\section{Age Sensitive Update Traffic over TCP}
\label{sec:TCP}
Before we delve into the problem of end-to-end age control, we demonstrate why TCP as a choice of transport protocol is unsuitable for age sensitive traffic. Specifically, we show that the congestion control mechanism of TCP together with its goal of guaranteed and ordered delivery of packets can lead to a very high age at the monitor, in comparison to when UDP is used, for a wide range of utilization of the network by the traffic generated by the source, and not just when the utilization is high.

We simulated a simple network consisting of a source that sends measurement updates to a monitor via a single Internet Protocol (IP) router. The source node has a bidirectional point-to-point (P2P) link of rate $1$ Mbps to the router. A similar link connects the router to the monitor. The source uses a TCP client to connect to a TCP server at the monitor and sends its update packets over the resulting TCP connection. We will also compare the obtained age with when UDP is used instead.

\emph{Retransmissions and In-order Delivery:} Figure~\ref{fig:UDPvsTCPPacketError} illustrates the impact of packet error on TCP. A packet was dropped independently of other packets with probability $0.1$. The figure compares the average age at the monitor and the average update packet delay, which is the time elapsed between generation of a packet at the source and its delivery at the monitor, when using TCP and UDP. On using TCP, the time average age achieves a minimum value of $0.18$ seconds when the source utilizes a fraction $0.2$ of the available $1$ Mbps to send update packets. This is clearly much larger than the minimum age of $\approx 0.01$ seconds at a utilization of $\approx 0.8$ when UDP is used. 

The large minimum age when using TCP is explained by the way TCP guarantees in order packet delivery to the receiving application (monitor). It causes fresher updates that have arrived out-of-order at the TCP receiver to wait for older updates that have not yet been received, for example, because of packet losses in the network. This can be seen in Figure~\ref{fig:UDPvsTCPPacketErrorRXBuffer} that shows how large measured packet delays coincide with a spike in the number of bytes received by the monitor application. The large delay is that of a received packet that had to undergo a TCP retransmission. The corresponding spike in received bytes, which is preceded by a pause, is because bytes with fresher information received earlier but out of order are held by the TCP receiver till the older packet is received post retransmission. Unlike TCP, UDP ignores dropped packets and delivers packets to applications as soon as they are received. This makes it desirable for age sensitive applications. As we will see later, ACP uses UDP to provide update packets with end-to-end transport over IP.

\emph{TCP Congestion Control and Small Packets:} Next, we describe the impact of small packets on the TCP congestion algorithm and its impact on age. This is especially relevant to a source sending measurement updates as the resulting packets may have small application payloads. Note that no packet errors were introduced in simulations used to make the following observations. Observe in the upper plot of Figure~\ref{fig:TCPSmallPacketSizes} that the $500$ byte packet payloads experience higher age at the monitor than the larger $536$ byte packets. The reason is explained by the impact of packet size on how quickly the size of the TCP congestion window (CWND) increases. The congestion window size doesn't increase till a sender maximum segment size (SMSS) bytes are acknowledged. TCP does this to optimize the overheads associated with sending payload. Packets with fewer bytes may thus require multiple TCP ACK(s) to be received for the congestion window to increase. This explains the slower increase in the size of the congestion window for $500$ byte payloads seen in Figure~\ref{fig:TCPSmallPacketSizes}. This causes smaller packets to wait longer in the TCP send buffer before they are sent out by the TCP sender, which explains the larger age in Figure~\ref{fig:TCPSmallPacketSizes}.
\section{The Age Control Protocol}
\label{sec:ACPProtocol}

\begin{figure}[!t]             
	\begin{center}
		\includegraphics[width=0.4\textwidth]{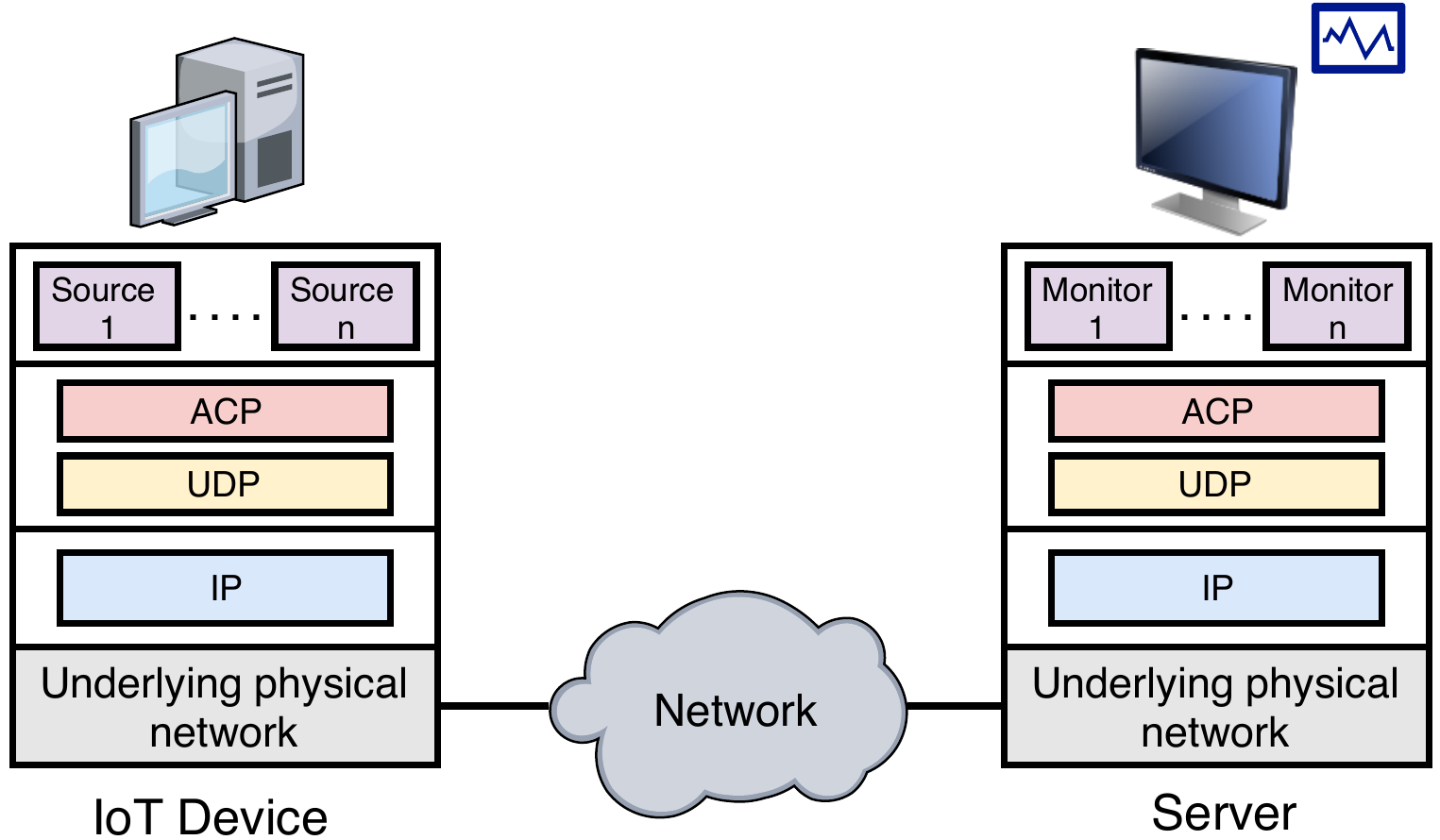}
		\caption{The ACP end-to-end connection.}
		\label{fig:networkStack}
	\end{center}
\vspace{-0.2in}
\end{figure}

The Age Control Protocol resides in the transport layer of the TCP/IP networking stack and operates only on the end hosts. Figure~\ref{fig:networkStack} shows an \emph{end-to-end connection} between two hosts, an IoT device, and a server, over the Internet. A source opens an ACP connection to its monitor. Multiple sources may connect to the same monitor. ACP uses the unreliable transport provided by the user datagram protocol (UDP) for sending of updates generated by the sources. This is in line with the requirements of fresh delivery of updates. Retransmissions make an update stale and also compete with fresh updates for network resources. 

The source ACP appends a header to an update from a source. The header contains a \emph{timestamp} field that stores the time the update was generated. The source ACP suggests to the source the rate at which it must generate updates. To be able to calculate the rate, the source ACP must estimate network conditions over the end-to-end path to the monitor ACP. This is achieved by having the monitor ACP acknowledge each update packet received from the source ACP by sending an ACK packet in return. The ACK contains the timestamp of the update being acknowledged. The ACK(s) allow the source ACP to keep an estimate of the age of sensed information at the monitor. An \emph{out-of-sequence} ACK, which is an ACK received after an ACK corresponding to a more recent update packet, is discarded by the source ACP. Similarly, an update that is received \emph{out-of-sequence} is discarded by the monitor. This is because the monitor has already received a more recent measurement from the source.

\begin{figure}[!t]             
	\begin{center}
		\includegraphics[width=0.47\textwidth]{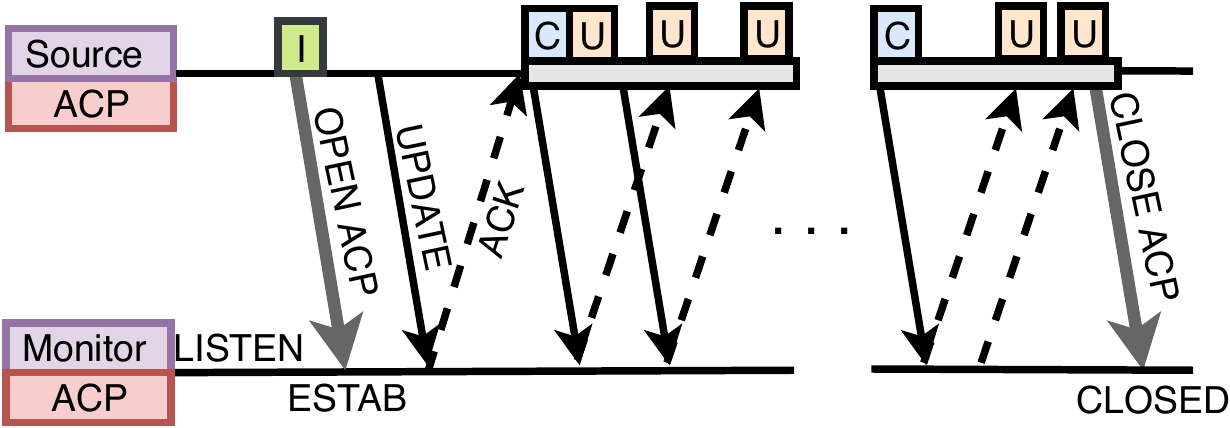}
		\caption{Timeline of an ACP connection. The box \textbf{I} marks the beginning of the initialization phase of ACP. The boxed \textbf{C} denotes the ACP algorithm (Algorithm~\ref{alg:acp}) executed when a new control epoch begins. The boxed \textbf{U} is executed when an ACK is received and updates $\overline{Z},\overline{\text{RTT}}, \text{and } \mathcal{T}$.}
		\label{fig:acpConnectionTimeline}
	\end{center}
\vspace{-0.2in}
\end{figure}

Figure~\ref{fig:acpConnectionTimeline} shows a timeline of a typical ACP connection. For an ACP connection to take place, the monitor ACP must be listening on a previously advertised UDP port. The ACP source first establishes a UDP connection with the monitor. This is followed by an \emph{initialization} phase during which the source sends an update and waits for an ACK or for a suitable timeout to occur, and repeats this process for a few times, with the goal of probing the network to set an initial update rate. Following this phase, the ACP connection may be described by a sequence of \emph{control epochs}. The end of the \emph{initialization} phase marks the start of the first control epoch. At the beginning of each control epoch, ACP sets the rate at which updates generated from the source are sent until the beginning of the next epoch. 


%

\section{The Age Control Problem}
\label{sec:problem}

\begin{figure}[!b]             
	\begin{center}
		\includegraphics[width=0.47\textwidth]{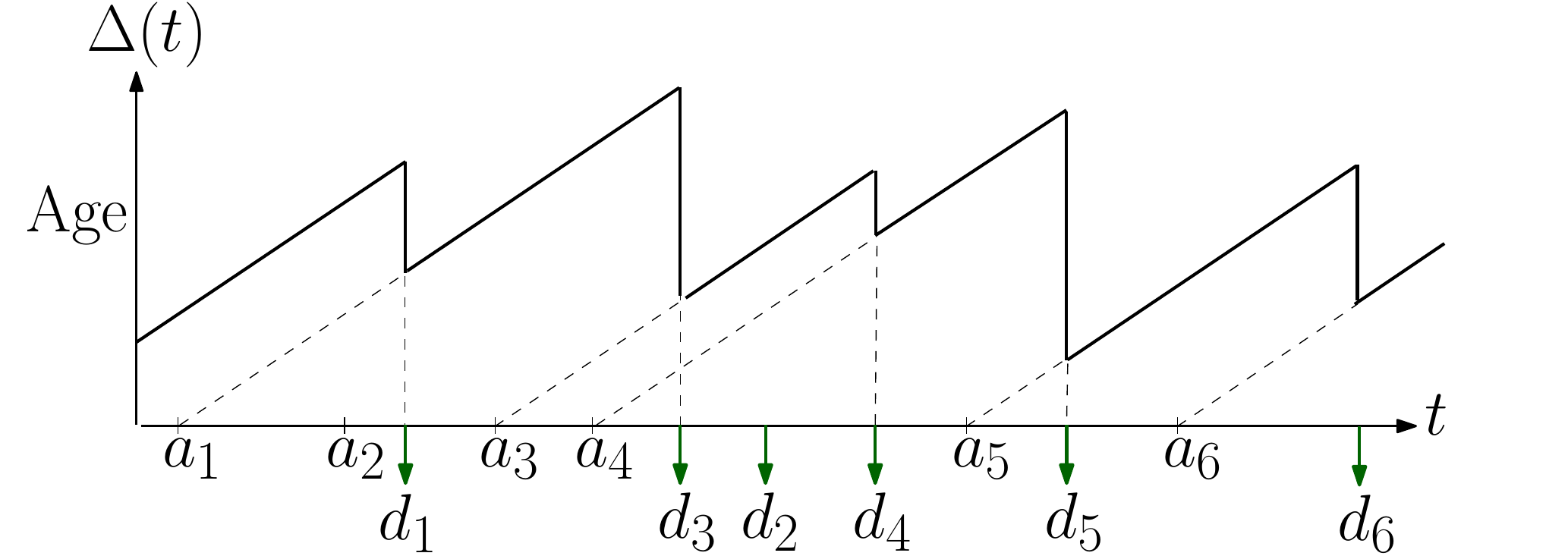}
		\caption{A sample function of the age $\age(t)$. Updates are indexed $1,2,\ldots$. The timestamp of update $i$ is $a_i$. The time at which update $i$ is received by the monitor is $d_i$. Since update $2$ is received out-of-sequence, it doesn't reset the age process.}
		\label{fig:ageSampleFunction}
	\end{center}
	\vspace{-0.2in}
\end{figure}

We will formally define the age of sensed information at a monitor. To simplify presentation, in this section, we will assume that the source and monitor are time synchronized, although the functioning of ACP doesn't require the same. Let $z(t)$ be the timestamp of the freshest update received by the monitor up to time $t$. Recall that this is the time the update was generated by the source.

The age at the monitor is $\age(t) = t - z(t)$ of the freshest update available at the monitor at time $t$. An example sample function of the age stochastic process is shown in Figure~\ref{fig:ageSampleFunction}. The figure shows the timestamps $a_1, a_2,\ldots,a_6$ of $6$ packets generated by the source. Packet $i$ is received by the monitor at time $d_i$. At time $d_i$, packet $i$ has age $d_i - a_i$. The age $\age(t)$ at the monitor increases linearly in between reception of updates received in the correct sequence. Specifically, it is reset to the age $d_i - a_i$ of packet $i$, in case packet $i$ is the freshest packet (one with the most recent timestamp) at the monitor at time $d_i$. For example, when update $3$ is received at the monitor, the only other update received by the monitor until then was update $1$. Since update $1$ was generated at time $a_1 < a_3$, the reception of $3$ resets the age to $d_3 - a_3$ at time $d_3$. On the other hand, while update $2$ was sent at a time $a_2 < a_3$, it is delivered out-of-order at a time $d_2 > d_3$. So packet $2$ is discarded by the monitor ACP and age stays unchanged at time $d_2$.

We want to choose the rate $\lambda$ (updates/second) that minimizes the expected value $\lim_{t \to \infty}E[\age(t)]$ of age at the monitor, where the expectation is over any randomness introduced by the network. Note that in the absence of a priori knowledge of a network model, as is the case with the end-to-end connection over which ACP runs, this expectation is unknown to both source and monitor and must be estimated using measurements. Lastly, we would like to dynamically adapt the rate $\lambda$ to nonstationarities in the network.

\begin{figure}[!b]             
	\begin{center}
		\includegraphics[width=0.45\textwidth]{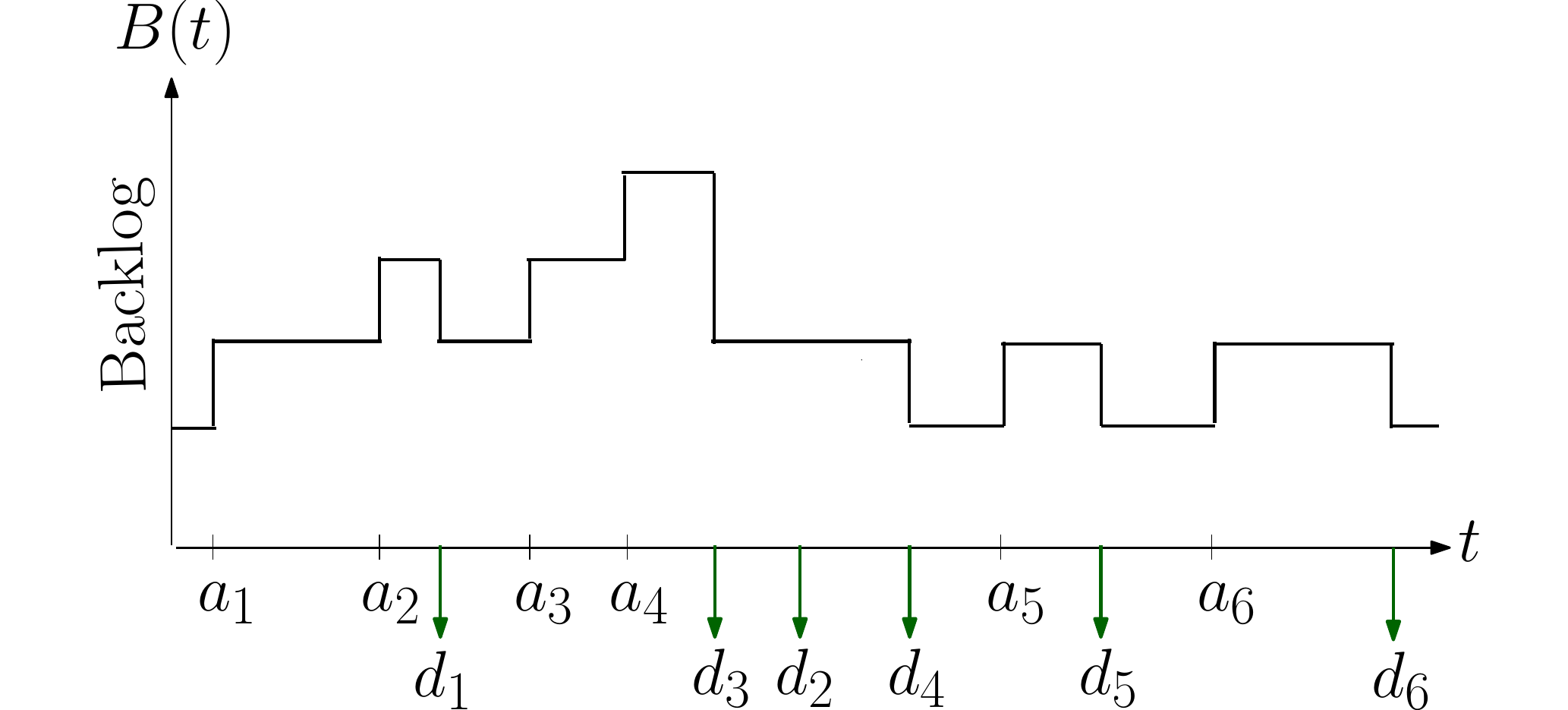}
		\caption{A sample function of the backlog process $B(t)$. Updates are indexed $1,2,\ldots$. The timestamp of update $i$ is $a_i$. The time at which update $i$ is received by the monitor is $d_i$. Since update $3$ is received before $2$, backlog is reduced by $2$ packets at $d_3$. Also, there is no change in $B(t)$ at $d_2 > d_3$.}
		\label{fig:backlogSampleFunction}
	\end{center}
\end{figure}

\begin{figure*}[!ht]             
	\begin{center}
		\subfloat[]{\includegraphics[width=.32\textwidth]{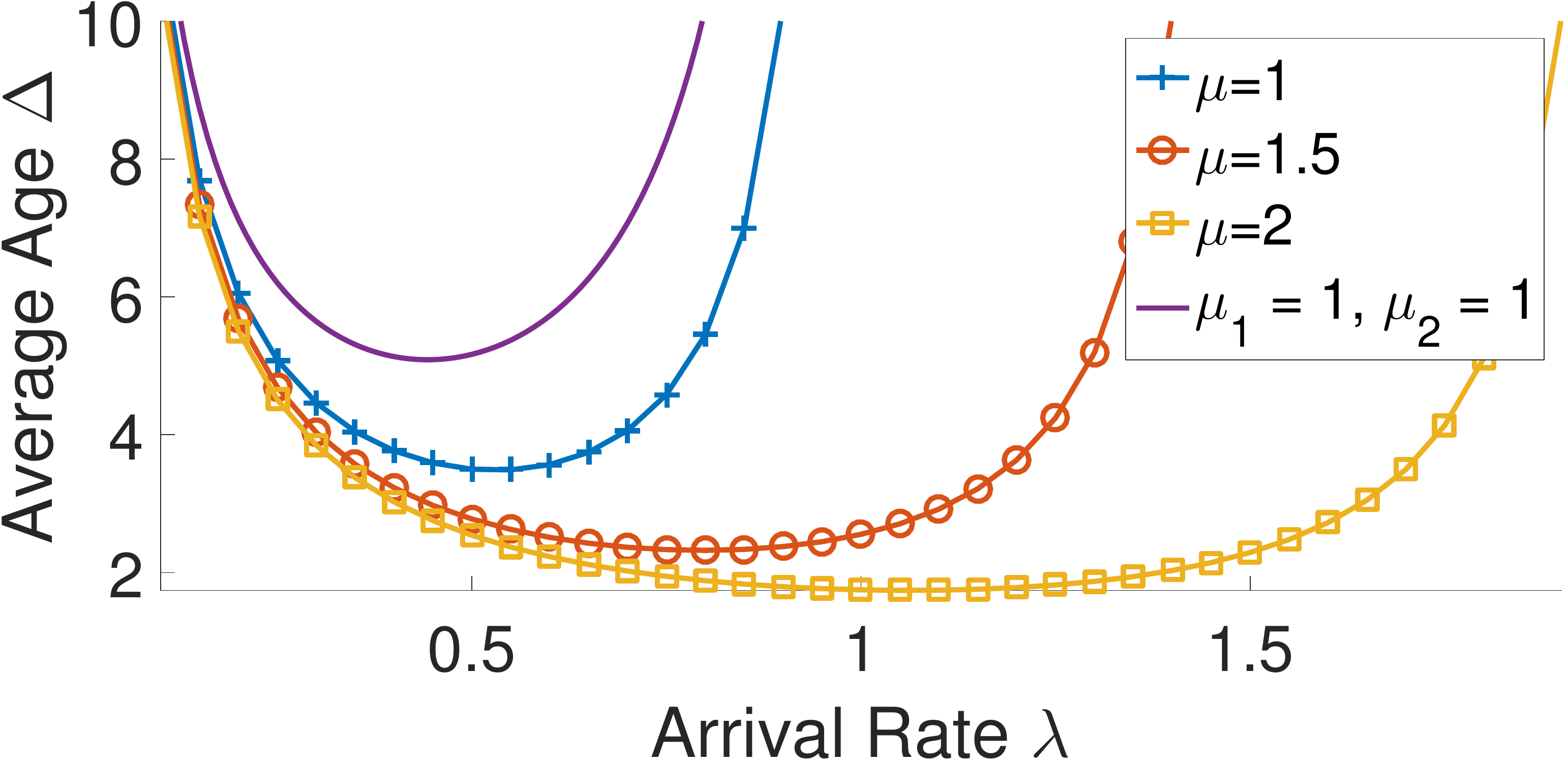}
			\label{fig:motivatingACP_age_lambda}}
		\hspace{.001in}
		\subfloat[]{\includegraphics[width=.32\textwidth]{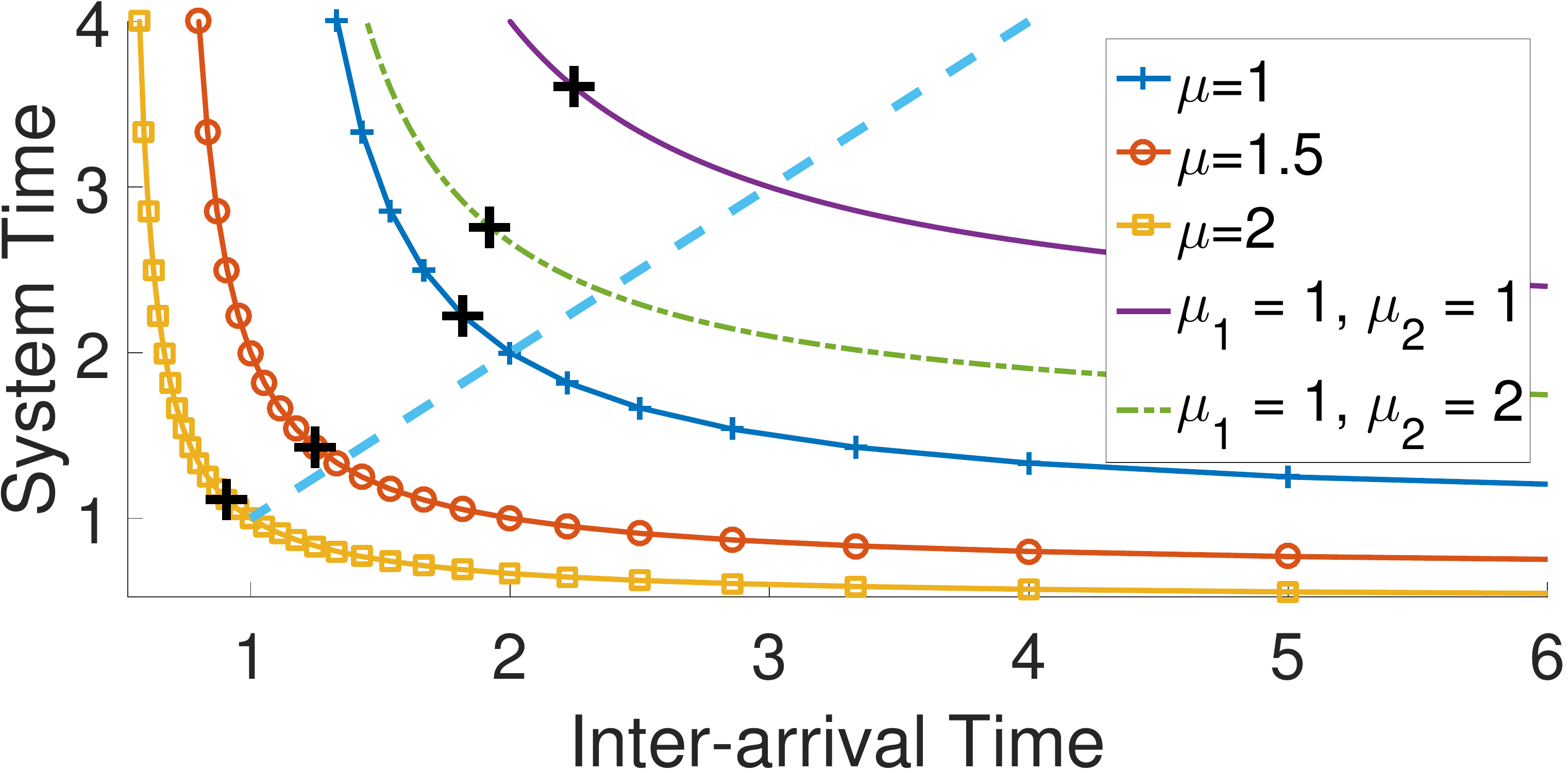}
			\label{fig:motivatingACPArrivalAndSystem}}
		\hspace{.001in}
		\subfloat[$\mu_1 = 1.$]{\includegraphics[width=.32\textwidth]{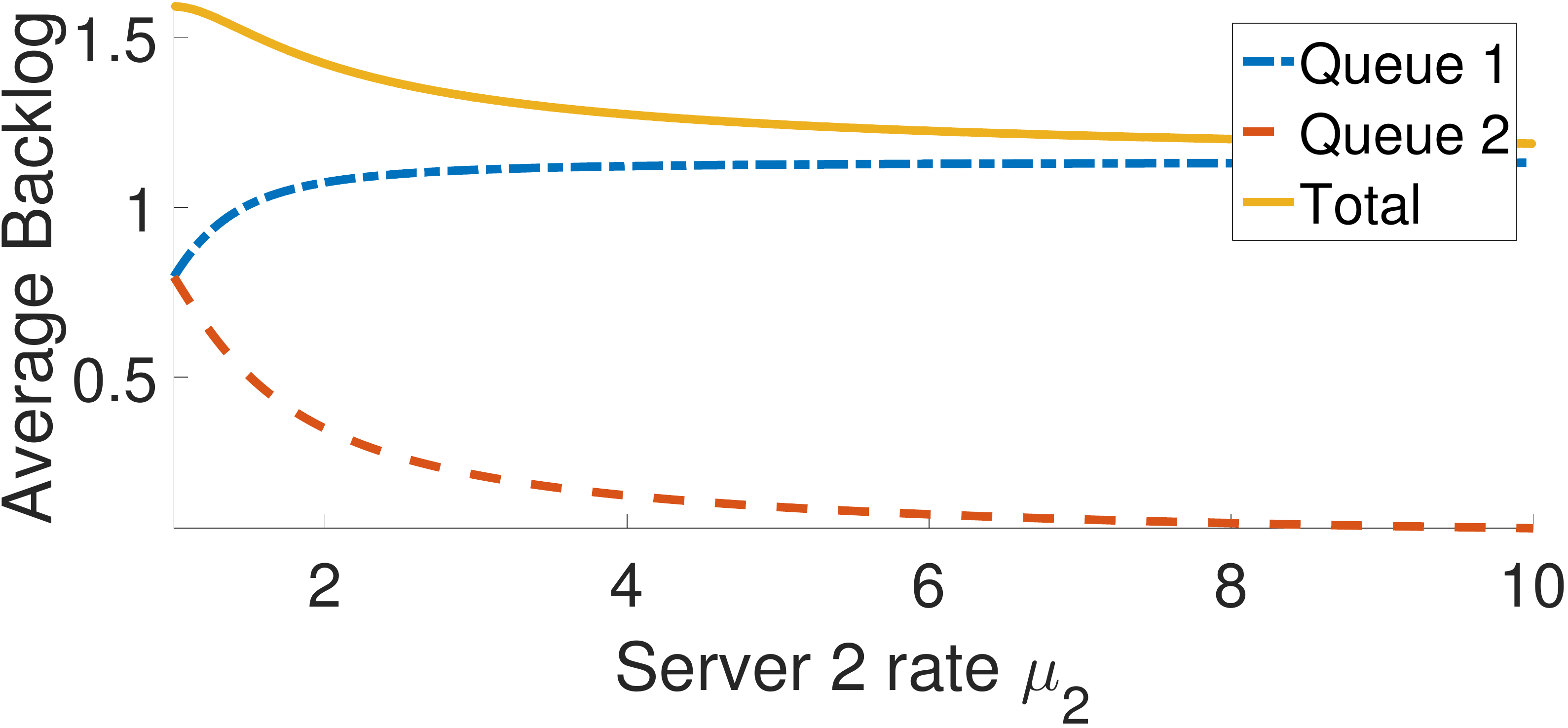}
			\label{fig:motivatingACPBacklog}}
		\caption{(a) Expected value of age as a function of update rate $\lambda$ is shown for different queueing networks. (b) Average update packet system time is shown as a function of inter-arrival time ($1/\lambda$). The green dashed line is the $45^{\circ}$ line. The $\mathrel +$ mark the age minimizing $\lambda$. (c) The backlog in each queue and the sum is shown as the service rate of the second queue increases from $1$ to $5$. Service rate of the first queue is $\mu_1 = 1$.}
		\label{fig:motivatingACP}
	\end{center}
\end{figure*}

\section{Good Age Control Behavior and Challenges}
\label{sec:acpIntuit}

ACP must suggest a rate $\lambda$ updates/second at which a source must send fresh updates to its monitor. ACP must adapt this rate to network conditions. To build intuition, let's suppose that the end-to-end connection is well described by an idealized setting that consists of a single FCFS queue that serves each update in constant time.  An update generated by the source enters the queue, waits for previously queued updates, and then enters service. The monitor receives an update once it completes service. Note that every update must age at least by the (constant) time it spends in service, before it is received by the monitor. It may age more if it ends up waiting for one or more other updates to complete service. 

In this idealized setting, one would want a new update to arrive as soon as the last generated update finishes service. To ensure that the age of each update received at the monitor is the minimum, one must choose a rate $\lambda$ such that new updates are generated in a periodic manner with the period set to the time an update spends in service. Also, update generation must be synchronized with service completion instants so that a new update enters the queue as soon as the last update finishes service. In fact, such a rate $\lambda$ is age minimizing even when updates pass through  a sequence of $Q>1$ such queues in tandem~\cite{shortle2018fundamentals}. The update is received by the monitor when it leaves the last queue in the sequence. The rate $\lambda$ will ensure that a generated packet ages exactly $Q$ times the time it spends in the server of any given queue. At any given time, there will be exactly $Q$ update packets in the network, one in each server. 

Of course, the assumed network is a gross idealization. We assumed a series of similar constant service facilities and that the time spent in service and instant of service completion were known exactly. We also assumed lack of any other traffic. However, as we will see further, the resulting intuition is significant. Specifically, \emph{a good age control algorithm must strive to have as many update packets in transit as  possible while simultaneously ensuring that these updates avoid waiting for other previously queued  updates}.

Before we detail our proposed control method, we will make a few salient observations using analytical results for simple queueing models and simulation results that capture stochastic service and generation of updates. These will help build on our intuition and also elucidate the challenges of age control over a priori unknown and likely non-stationary end-to-end network conditions. 

\subsection{Analytical Queueing Model for Two Queues} We will consider two queueing models. One is the $M/M/1$ FCFS queue with an infinite buffer in which a source sends update packets at a rate $\lambda$ to a monitor via a single queue, which services packets at a rate $\mu$ updates per second. The updates are generated as a Poisson process of rate $\lambda$ and packet service times are exponentially distributed with $1/\mu$ as the average time it takes to service a packet. In the other model, updates travel through two queues in tandem. Specifically, they enter the first queue that is serviced at the rate $\mu_1$. On finishing service in the first queue, they enter the second queue that services packets at a rate of $\mu_2$. As before, updates arrive to the first queue as a Poisson process and packet service times are exponentially distributed. The average age for the case of a single $M/M/1$ queue was analyzed in~\cite{KaulYatesGruteser-Infocom2012}. We extend their analysis to obtain analytical expressions of average age as a function of $\lambda,\mu_1$ and $\mu_2$ for the two queue case, by using the well known result that updates also enter the second queue as a Poisson process of rate $\lambda$~\cite{shortle2018fundamentals}.

\emph{On the impact of non-stationarity and transient network conditions:} Figure~\ref{fig:motivatingACP_age_lambda} shows the expected value (average) of age as a function of $\lambda$ when the queueing systems are in \emph{steady state}. It is shown for three single $M/M/1$ queues, each with a different service rate, and for two queues in tandem with both servers having the same unit service rate. Observe that all the age curves have a bowl-like shape that captures the fact that a too small or a too large $\lambda$ leads to large age. Such behavior has been observed in non-preemptive queueing disciplines in which updates can't preempt other older updates. A reasonable strategy to find the optimal rate thus seems to be one that starts at a certain initial $\lambda$ and changes $\lambda$ in a direction such that a smaller expected age is achieved. 

In practice, the absence of a network model (unknown service distributions and expectations), would require Monte-Carlo estimates of the expected value of age for every choice of $\lambda$. Getting these estimates, however, would require averaging over a large number of instantaneous age samples and would slow down adaptation. This could lead to updates experiencing excessive waiting times when $\lambda$ is too large. Worse, transient network conditions (a run of bad luck) and non-stationarities, for example, because of introduction of other traffic flows, could push these delays to even higher values, leading to an even larger backlog of packets in in transit. Figure~\ref{fig:motivatingACP_age_lambda}, illustrates how changes in network conditions (service rate $\mu$ and number of hops (queues)) can lead to large changes in the expected age. 

It is desirable for a good age control algorithm to not allow the end-to-end connection to drift into a high backlog state. As we describe in the next section, ACP tracks changes in the average number of backlogged packets and average age over short intervals, and in case backlog and age increase, ACP acts to rapidly reduce the backlog.

\emph{On Optimal Average Backlogs:} Figure~\ref{fig:motivatingACPArrivalAndSystem} plots the average packet system times, where the system time of a packet is the time that elapses between its arrival and completion of its service, as a function of inter-arrival time ($1/\lambda$) for three single queue $M/M/1$ networks and two networks that have two queues in tandem. As expected, increase in inter-arrival time reduces the system time. As inter-arrival times become large, packets wait less often for others to complete service. As a result, as inter-arrival time increases, the system times converge to the average service time of a packet. For each queueing system, we also mark on its plot the inter-arrival time that minimizes age. It is instructive to note that for the three single queue systems this inter-arrival time is only slightly smaller than the system time. However, for the two queues in tandem with service rates of $1$ each, the inter-arrival time is a lot smaller than the system time. The implication being that on an average it is optimal to send slightly more than one ($\approx 1.2$) packet every system time for the single queue system. However, for the two queue network with the similar servers, we want to send a larger number ($\approx 1.6$) of packets every system time. For the two queue network where the second queue is served by a faster server, this number is smaller ($\approx 1.43$). As we observe next, as one of the servers becomes faster, the two queue network becomes more akin to a single queue network with the slower server.

\begin{table}
	\begin{center}
		\begin{tabular}[!b]{|L|L|L|L|L|L|L|}
			\hline
			\text{Network} & R_1 & R_2 & R_3 & R_4 & R_5 & R_6\\
			\hline
			\text{Net A} & 1 & 1 & 1 & 1 & 1 & 1\\
			\hline
			\text{Net B} & 1 & 1 & \mathbf{5} & \mathbf{5} & 1 & 1\\
			\hline
			\text{Net C} & 1 & \mathbf{5} & 5 & 5 & \mathbf{5} & 1\\
			\hline
			\text{Net D} & \mathbf{5} & 5 & 5 & 5 & 5 & 1\\
			\hline	
			\text{Net E} & 5 & 5 & 5 & 5 & 5 & \mathbf{5}\\
			\hline
		\end{tabular}
		\caption{Various P2P link configurations applied to the network diagram in Figure~\ref{fig:simulationNetwork}. The rates $R_i$ are in Mbps. $R_1$ is the rate of the link between the source and AP-1 and $R_6$ is that of the link between AP-2 and the monitor.}
		\label{tab:goodACPSimulationTable}
	\end{center}
\vspace{-0.2in}
\end{table}

Note that these numbers are in fact the optimal (age minimizing) average number of packets in the system. Figure~\ref{fig:motivatingACPBacklog} shows how this optimal average backlog varies as a function of $\mu_2$ for a given $\mu_1$. The observations stay the same on swapping $\mu_1$ and $\mu_2$. As $\mu_2$ increases, that is as the second server becomes faster than the first ($\mu_1 = 1$), we see that the average backlog increases in queue $1$ and reduces in queue $2$, while the sum backlog gets closer to the optimal backlog for the single queue case. Specifically, as queue $1$ becomes a larger bottleneck relative to queue $2$, optimal $\lambda$ must adapt to the bottleneck queue. The backlog in the faster queue is governed by the resulting choice of $\lambda$. For when the rates $\mu_1$ and $\mu_2$ are similar, they see similar backlogs. However, as is seen for when $\mu_1,\mu_2 = 1$, the backlog per queue is smaller than a network with only a single such queue. However, the sum backlog ($\approx 1.6$) is larger.

\begin{figure}[!t]             
	\begin{center}
			\includegraphics[width=.5\textwidth]{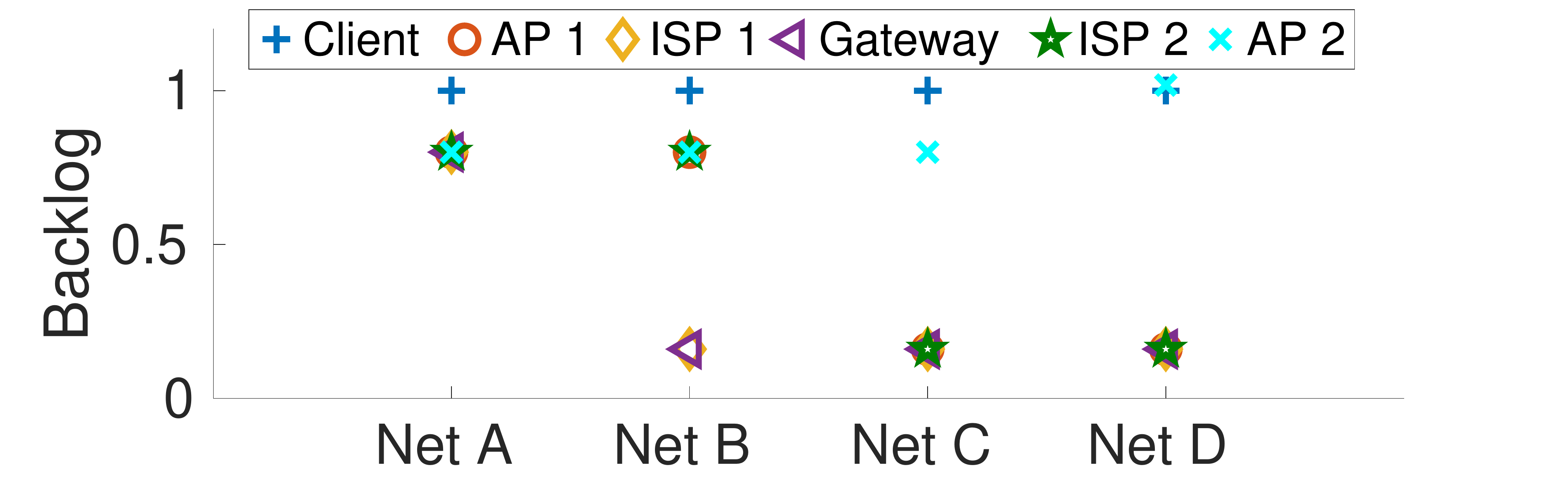}
			\caption{Average backlogs at different nodes in the network, shown in Fig~\ref{fig:simulationNetwork}, at the optimal update rate. Net E is similar to Net A and not shown.}
			\label{fig:motivatingACPBacklog}
	\end{center}
\vspace{-0.1in}
\end{figure}

\subsection{Simulating Larger Number of Hops} 
\begin{figure}[!t]             
	\begin{center}
		\includegraphics[width=0.48\textwidth]{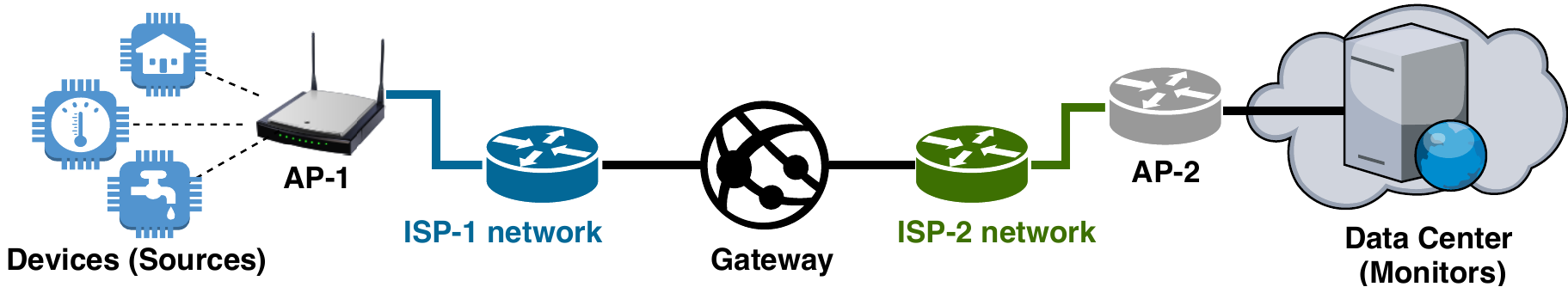}
		\caption {Sources are connected to the monitor via multiple routers and access points. Each source update travels over six hops. The first hop is between the source and access point AP-1. This could be either P2P or WiFi. The other hops that involve the ISP(s) and the Gateway are an abstraction of the Internet. These hops are P2P links and we vary their rates to simulate different end-to-end RTT.}
		\label{fig:simulationNetwork}
	\end{center}
\vspace{-0.2in}
\end{figure}

To see if this intuition generalizes to more number of hops, we simulated an end-to-end connection which has the source send its packet to the monitor over $6$ hops, where each hop is serviced by a bidirectional P2P link. The hops are shown in Figure~\ref{fig:simulationNetwork}. We vary the rates at which the P2P links transmit packets to gain insight into how queues in a network must be populated with update packets at an age optimal rate. We also introduce other traffic in the network that occupies, on an average, a fraction $0.2$ Mbps of each P2P link from the source to the monitor. The different configurations are summarized in Table~\ref{tab:goodACPSimulationTable}. For each network configuration we have the source send updates over UDP to the monitor using an a priori chosen rate $\lambda$. We vary $\lambda$ over a wide range of values and for each $\lambda$ we calculate the obtained time average age. These simulations allow us to empirically pick the age minimizing $\lambda$ for the given network. 

Figure~\ref{fig:motivatingACPBacklog} shows the time average \emph{backlog} (queue occupancy) at the different nodes in the network at the optimal $\lambda$. The backlog at a node includes the update packet being transmitted on a node's outgoing P2P link and any update that is awaiting transmission at the node. Observe that all P2P links in each of Net A and Net E have the same rate, $1$ and $5$ Mbps respectively. Though Net B has links much faster than that of Net A, for both these networks the average backlog at all nodes is close to $1$. That it is smaller than $1$ is explained by the presence of the other flow. The other flow, which also originates at the client, is also the reason why the client sees a slightly larger average queue occupancy by the update packets.

Net B has faster P2P links connecting ISP(s) and the Gateway when compared to Net A. However, its other links are slower than that in Net E. We see that the nodes that have fast outgoing links have low backlogs and those that have slow links have an average backlog close to $0.8$. The source has a slow outgoing link and as a result of the other flow sees slightly larger occupancy of update packets. In summary, at $\lambda$ that minimizes average age, as is also shown in Figure~\ref{fig:motivatingACPBacklog} for Net C and Net D, \emph{nodes with outgoing links that are bottlenecks relative to the others' links see a backlog such that no more than one update packet is queued at them}. Naturally, nodes with faster links see smaller backlogs in proportion to how fast their links are with respect to the bottleneck.

A corollary to the above observations, which we do not demonstrate for lack of space, is that a good age control algorithm should on an average have a larger number of packets simultaneously in transit in a network with a larger number of hops (nodes/queues).

A good age control algorithm must not allow the end-to-end connection to drift into a high backlog state. As described next, ACP tracks changes in the number of backlogged packets and average age over short intervals, and in case backlog and age increase, ACP acts to rapidly reduce the backlog.

\section{The ACP Control Algorithm}
\label{sec:algorithm}

\begin{figure}[!t]             
	\begin{center}
		\includegraphics[width=0.45\textwidth]{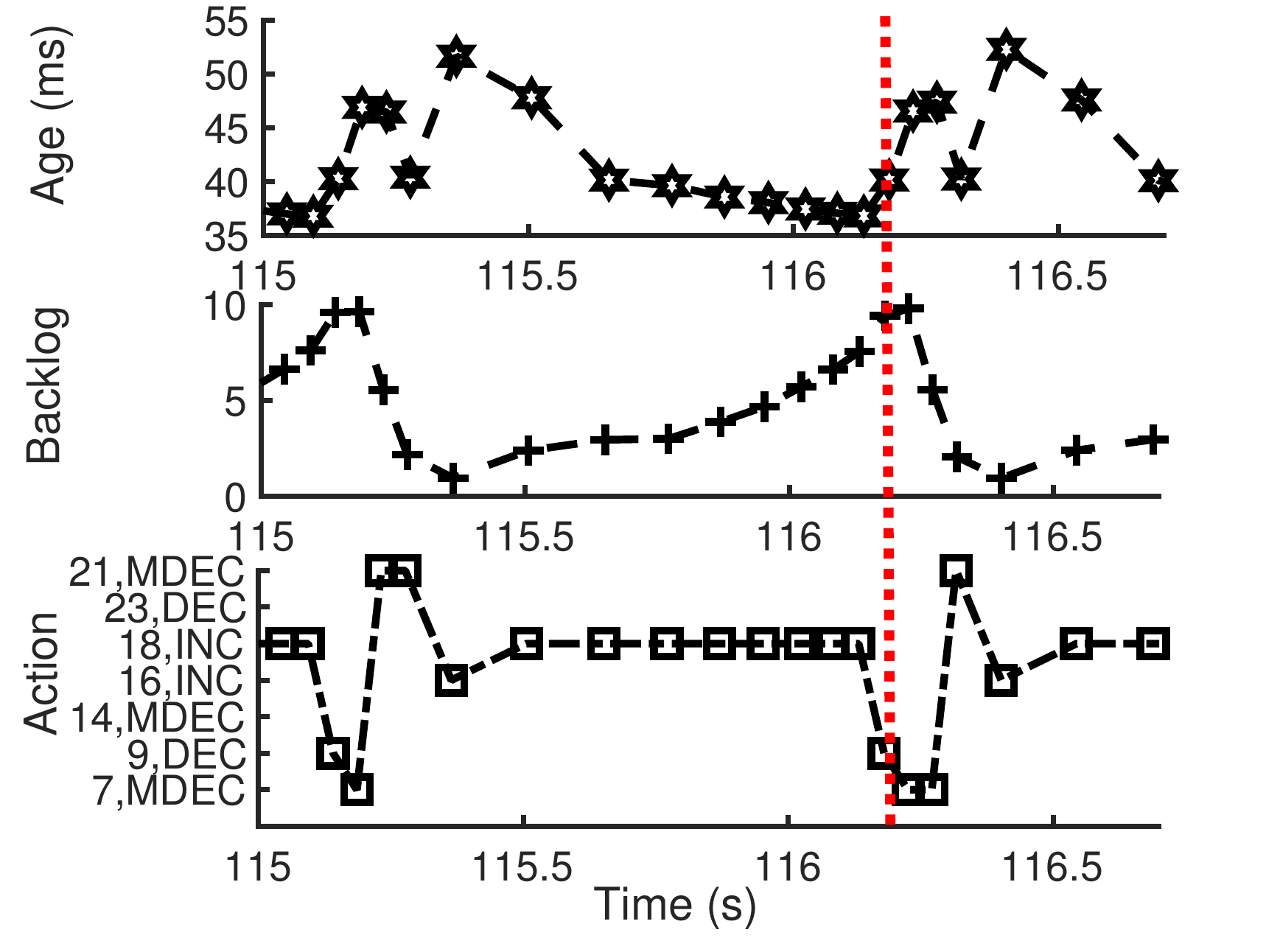}
		\caption {A snippet from the function of ACP. The y-axis of the plot showing actions denotes the action and the line number in Algorithm~\ref{alg:acp}. Note the action marked by the dotted red line. At the time instant ACP observes an increase in both backlog and age and chooses ($9$,DEC) initially. However, there is still a significant jump in age. This results in the choice of multiplicative decrease ($7$,MDEC).}
		\label{fig:acpDemo}
	\end{center}
	\vspace{-0.2in}
\end{figure}

Let the control epochs of ACP (Section~\ref{sec:problem}) be indexed $1,2,\ldots$. Epoch $k$ starts at time $t_k$. At $t_1$ the update rate $\lambda_{1}$ is set to the inverse of the average packet round-trip-times (RTT) obtained at the end of the initialization phase. At time $t_k$, $k>1$, the update rate is set to $\lambda_k$. The source transmits updates at a fixed period of $1/\lambda_{k}$ in the interval $(t_{k}, t_{k+1})$.

Let $\overline{\age}_k$ be the estimate at the source ACP of the time average update age at the monitor at time $t_k$. This average is calculated over $(t_{k-1}, t_k)$. To calculate it, the source ACP must construct its estimate of the age sample function (see Figure~\ref{fig:ageSampleFunction}), over the interval, at the monitor. It knows the time $a_i$ a source sent a certain update $i$. However, it needs the time $d_i$ at which update $i$ was received by the monitor, which it approximates by the time the ACK for packet $i$ was received. On receiving the ACK, it resets its estimate of age to the resulting round-trip-time (RTT) of packet $i$. 

Note that this value is an overestimate of the age of the update packet when it was received at the monitor, since it includes the time taken to send the ACK over the network. The time average $\overline{\age}_k$ is obtained simply by calculating the area under the resulting age curve over $(t_{k-1}, t_k)$ and dividing it by the length $t_k - t_{k-1}$ of the interval.

Let $\overline{B}_k$ be the time average of backlog calculated over the interval $(t_{k-1}, t_k)$. This is the time average of the instantaneous backlog $B(t)$ over the interval. The instantaneous backlog increases by $1$ when the source sends a new update. When an ACK corresponding to an update $i$ is received, update $i$ and any unacknowledged updates older than $i$ are removed from the instantaneous backlog. 

In addition to using RTT(s) of updates for age estimation, we also use them to maintain an exponentially weighted moving average (EWMA) $\overline{\text{RTT}}$ of RTT. We update $\overline{\text{RTT}} = (1 - \alpha) \overline{\text{RTT}} + \alpha \text{RTT}$ on reception of an ACK that corresponds to a round-trip-time of RTT. 
\begin{algorithm}[t]
\caption{Control Algorithm of ACP}
\label{alg:acp}
\footnotesize
\begin{algorithmic}[1]
\State \textbf{INPUT:} $b_k, \delta_k, \overline{T}$
\State \textbf{INIT:} $flag \gets 0$, $\gamma \gets 0$
\While{true} 
	\If {$b_k>0$  \&\& $\delta_k >0$}\label{alg:one}
		 \If {$flag==1$}
		 \State $\gamma=\gamma+1$\label{alg:oneincr}
			\State MDEC($\gamma$)
		\Else
			\State DEC\label{alg:oneDEC}
		\EndIf
		\State $flag\gets 1$
	\ElsIf { $b_k>0$  \&\& $\delta_k <0$}\label{alg:two}
		\If {$flag==1$ \&\& $|b_k|< 0.5*|b_k^{*}|$}\label{alg:two1}
			\State $\gamma=\gamma+1$
			\State MDEC($\gamma$)
		\Else
			\State INC, $flag\gets 0$, $\gamma\gets0$ \label{alg:two2}
		\EndIf
	\ElsIf { $b_k<0$  \&\& $\delta_k >0$}
		\State INC, $flag\gets 0$, $\gamma\gets0$ \label{alg:three}
	\Else { $b_k<0$  \&\& $\delta_k <0$}\label{alg:four}
		 \If {$flag==1$ \&\& $\gamma>0$}
			\State MDEC($\gamma$) \label{alg:fourMDEC}
		\Else
			\State DEC, $flag\gets 0$, $\gamma\gets0$
		\EndIf
	\EndIf
\State update $\lambda_k$
\State wait $\overline{T}$ 	
\EndWhile
\end{algorithmic}
\end{algorithm}

The source ACP also estimates the inter-update arrival times at the monitor and the corresponding EWMA $\overline{Z}$. The inter-update arrival times are approximated by the corresponding inter-ACK arrival times. The length $\overline{T}$ of a control epoch is set as an integral multiple of $\mathcal{T} = \min(\overline{\text{RTT}}, \overline{Z})$. This ensures that the length of a control epoch is never too large and allows for fast enough adaptation. Note that at sufficiently low rate $\lambda_{k}$ of sending updates $\overline{Z}$ is large and at a sufficiently high update rate $\overline{\text{RTT}}$ is large. At time $t_k$ we set $t_{k+1} = t_{k} + \overline{T}$. In all our evaluation we have used $\overline{T} = 10 \mathcal{T}$. The resulting length of $\overline{T}$ was observed to be long enough to see desired changes in average backlog and age in response to a choice of source update rate at the beginning of an epoch. The source updates $\overline{\text{RTT}}$, $\overline{Z}$, and $\overline{T}$ every time an ACK is received.

At the beginning of control epoch $k > 1$, at time $t_k$, the source ACP calculates the difference $\delta_k = \overline{\age}_k - \overline{\age}_{k-1}$ in average age measured over intervals $(t_{k-1}, t_k)$ and $(t_{k-1}, t_{k-2})$ respectively. Similarly, it calculates $b_k = \overline{B}_k - \overline{B}_{k-1}$.

ACP at the source chooses an action $u_k$ at the $k$\textsuperscript{th} epoch that targets a change $b^{*}_{k+1}$ in average backlog over an interval of length $\mathcal{T}$ with respect to the $k$\textsuperscript{th} interval. The actions, may be broadly classified into (a) additive increase (INC), additive decrease (DEC), and multiplicative decrease (MDEC). MDEC corresponds to a set of actions $\text{MDEC}(\gamma)$, where $\gamma = 1,2,\ldots$. We have%
{\small
\begin{align}
&\text{INC: }  b^{*}_{k+1} = \kappa,\,
\text{DEC: }  b^{*}_{k+1} = -\kappa,\nonumber\\
&\text{MDEC(}\gamma{\text{): } }  b^{*}_{k+1} = -(1 - 2^{-\gamma}) B_k,
\end{align}
}
where $\kappa > 0$ is a step size parameter. 

ACP attempts to achieve $b^{*}_{k+1}$ by setting $\lambda_k$ appropriately. The estimate of $\overline{Z}$ at the source ACP of the average inter-update arrival time at the monitor gives us the rate $1/\overline{Z}$ at which updates sent by the source arrive at the monitor. This and $\lambda_k$ allow us to estimate the average change in backlog over $\mathcal{T}$ as $(\lambda_k - (1/\overline{Z})) \mathcal{T}$. Therefore, to achieve a change of $b^{*}_{k+1}$ requires choosing $\lambda_k = \frac{1}{\overline{Z}} + \frac{b^{*}_{k+1}}{\mathcal{T}}$. 
Algorithm~\ref{alg:acp} summarizes how ACP chooses its action $u_k$ as a function of $b_k$ and $\delta_k$. Figure~\ref{fig:acpDemo} shows an example of ACP in action.

\begin{figure}[!t]             
\begin{center}
\includegraphics[width=0.3\textwidth]{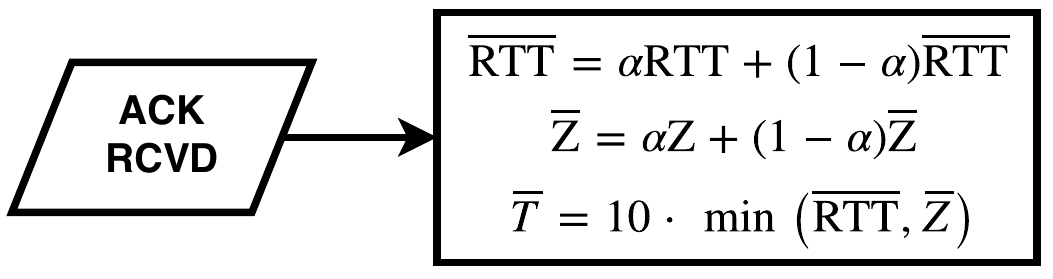}
\caption{Update of $\overline{\text{RTT}}$, $\overline{Z}$, and $\overline{T}$, which takes place every time an ACK is received.}
\label{fig:flow}
\end{center}
\end{figure}

The source ACP targets a reduction in average backlog over the next control interval in case either $b_k>0, \delta_k>0$ or $b_k<0, \delta_k<0$. The first condition (line~\ref{alg:one}) indicates that the update rate is such that updates are experiencing larger than optimal delays. ACP attempts to reduce the backlog, first using DEC (line~\ref{alg:oneDEC}), followed by multiplicative reduction MDEC to reduce congestion delays and in the process reduce age quickly. Consecutive occurrences ($flag == 1$) of this case (tracked by increasing $\gamma$ by $1$ in line~\ref{alg:oneincr}) attempt to decrease backlog even more aggressively, by a larger power of $2$.

The condition $b_k<0, \delta_k<0$ occurs on a reduction in both age and backlog. ACP greedily aims at reducing backlog further hoping that age will reduce too. It attempts MDEC (line~\ref{alg:fourMDEC}) if previously the condition $b_k>0, \delta_k>0$ was satisfied. Else, it attempts an additive decrease DEC.

The source ACP targets an increase in average backlog over the next control interval in case either $b_k>0, \delta_k<0$ or $b_k<0, \delta_k>0$. On the occurrence of the first condition (line~\ref{alg:three}) ACP greedily attempts to increase backlog.

When the condition $b_k<0, \delta_k>0$ occurs, we check if the previous action attempted to reduce the backlog. If not, it hints at too low an update rate causing an increase in age. So, ACP attempts an additive increase (line~\ref{alg:two2}) of backlog. If yes, and if the actual change in backlog was much smaller than the desired (line~\ref{alg:two1}), ACP attempts to reduce backlog multiplicatively. This helps counter situations where the increase in age is in fact because of increasing congestion. Specifically, increasing congestion in the network may cause the inter-update arrival rate $1/\overline{Z}$ at the monitor to reduce during the epoch. As a result, despite the attempted multiplicative decrease in backlog, it may change very little. Clearly, in such a situation, even if the backlog reduced a little, the increase in age was not caused because the backlog was low. The above check ensures ACP attempts reducing backlog to desired levels. In the above case, if instead ACP ignores the much smaller than desired change, it will end up increasing the rate of updates, further increasing backlog and age.

\section{Evaluation Methodology}
\label{sec:evaluation}


We used a mix of simulations and real-world experiments to evaluate ACP. While simulations allowed us to test with large numbers of sources contending with each other over wireless access under varied wireless channel conditions and densities of source placements, real-world experiments allowed us to test ACP over a real intercontinental end-to-end connection. 

Figure~\ref{fig:simulationNetwork} shows the end-to-end network used for simulations. We start by describing the wireless access over which sources connect to AP-1. We performed simulations for $1-50$ sources accessing AP-1 using the WiFi ($802.11$g) medium access. We simulated for sources spread uniformly and randomly over areas of $10\times 10$ m$^2$, $20\times 20$ m$^2$ and $50\times 50$ m$^2$. The channel between a source and AP-1 was chosen to be Log-Normally distributed with choices of $4$, $8$, and $12$ for the standard deviation. The pathloss exponent was $3$. WiFi physical (PHY) layer rates were set to one of $12$ Mbps and $54$ Mbps. We simulated for no WiFi retries and a max retry limit of $7$.

For the network beyond AP-1, all links were configured to be P2P. We set the P2P link rates from the set $\{0.3, 0.6, 1.2, 6.0\}$ Mbps. This was to simulate network RTT of a wide range.  We used the network simulator ns3\footnote{\url{https://www.nsnam.org/}} together with the YansWiFiPhyHelper\footnote{\url{https://www.nsnam.org/doxygen/classns3_1_1_yans_wifi_phy.html}}. Our simulated network is however limited in the number of hops, which is six.

We also evaluated ACP in the real-world by making $2 - 10$ sources connected to an enterprise WiFi access point, which is part of a university network, send their updates over the Internet to monitors that were running on a server with a global IP on another continent. This setup allows us to test ACP over a path with large RTT(s) and tens of hops. While the WiFi access point doesn't see much other traffic, we don't control the interference that may be created by adjoining access points or WiFi clients. Lastly, we had no control over the traffic on the university intranet when the experiments were performed.

To compare the age control performance of ACP, we use \emph{Lazy}. \emph{Lazy}, like ACP, also adapts the update rate to network conditions. However, it is very conservative and keeps the average number of update packets in transit small. Specifically, it updates the $\overline{\text{RTT}}$ every time an ACK is received and sets the current update rate to the inverse of $\overline{\text{RTT}}$. Thus, it aims at maintaining an average backlog of $1$.

We end by stating that an appropriate selection of step size $\kappa$ is crucial to the proper functioning of ACP. We chose it by hit and trial. For simulations, we found a step size of $\kappa = 0.25$ to be the best. However, this turned out to be too small for experiments over the Internet. For these, we tried $\kappa \in \{1, 2\}$.

Next, we will discuss the simulation results followed by the real-world results.

\section{Simulation Results}
\label{sec:simulationResults}
\begin{figure}[!t]             
	\begin{center}
		\subfloat[]{\includegraphics[width=.25\textwidth]{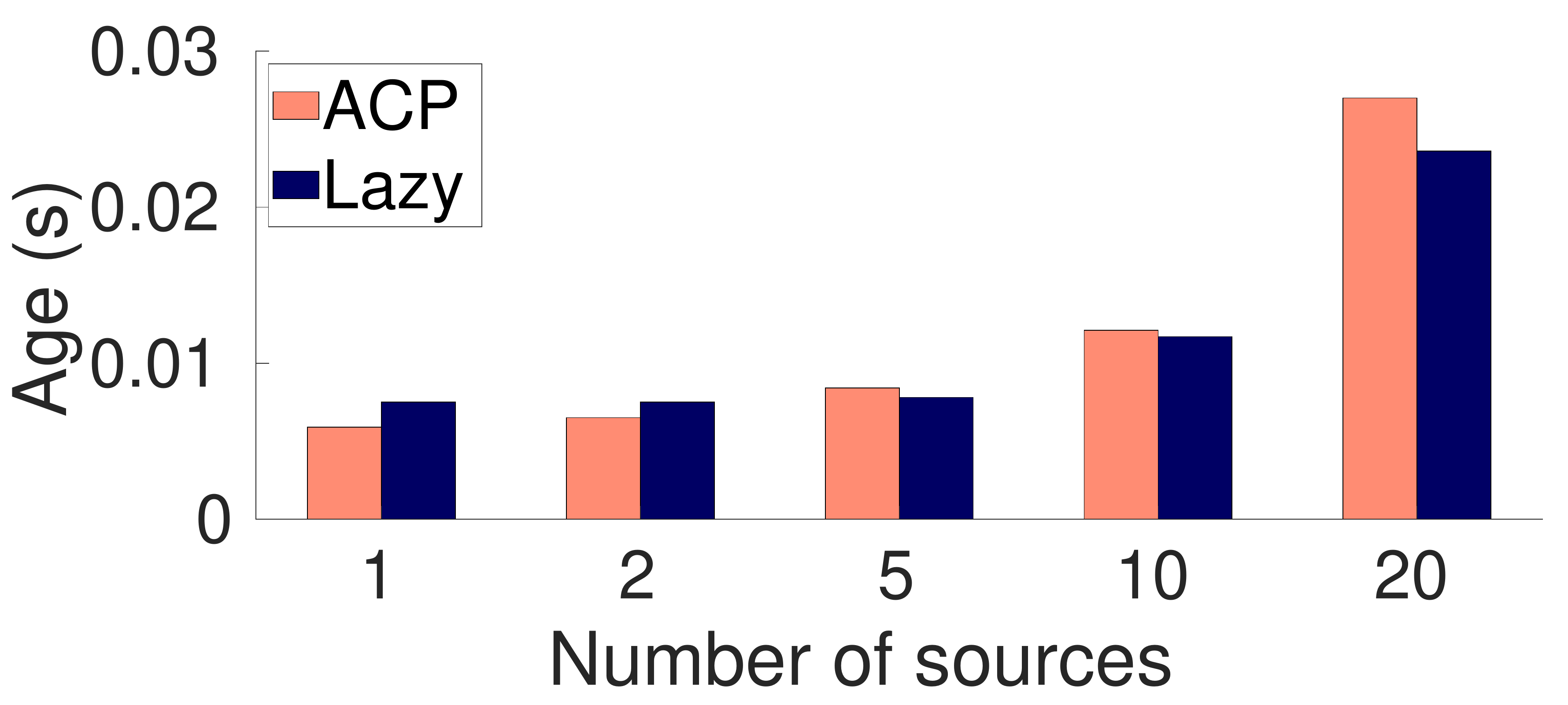}
			\label{fig:age_topology_1_fast}}
		\subfloat[]{\includegraphics[width=.25\textwidth]{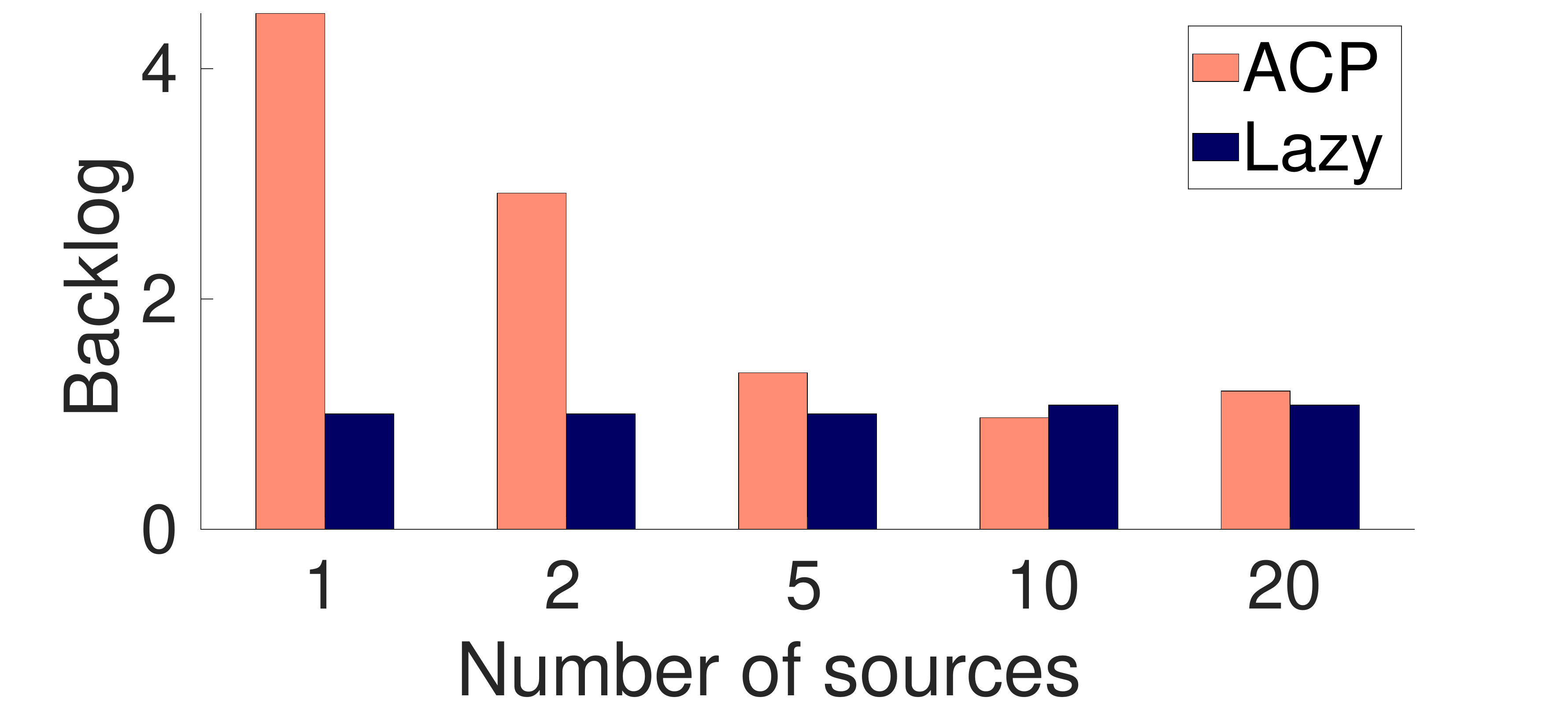}
			\label{fig:backlog_topology_1_fast}}\\
			\vspace{-0.02\textwidth}
				\subfloat[]{\includegraphics[width=.25\textwidth]{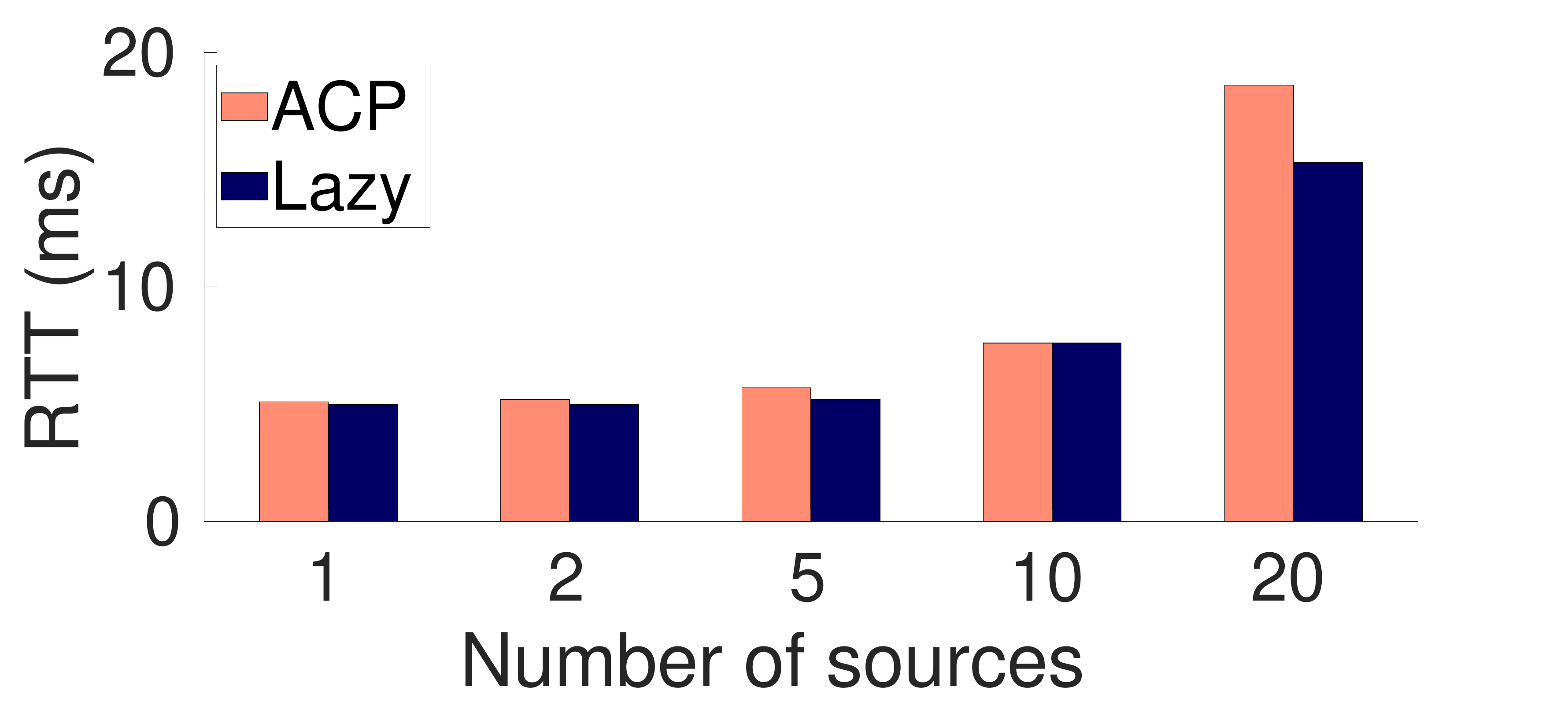}
			\label{fig:rtt_topology_1_fast}}
				\subfloat[]{\includegraphics[width=.25\textwidth]{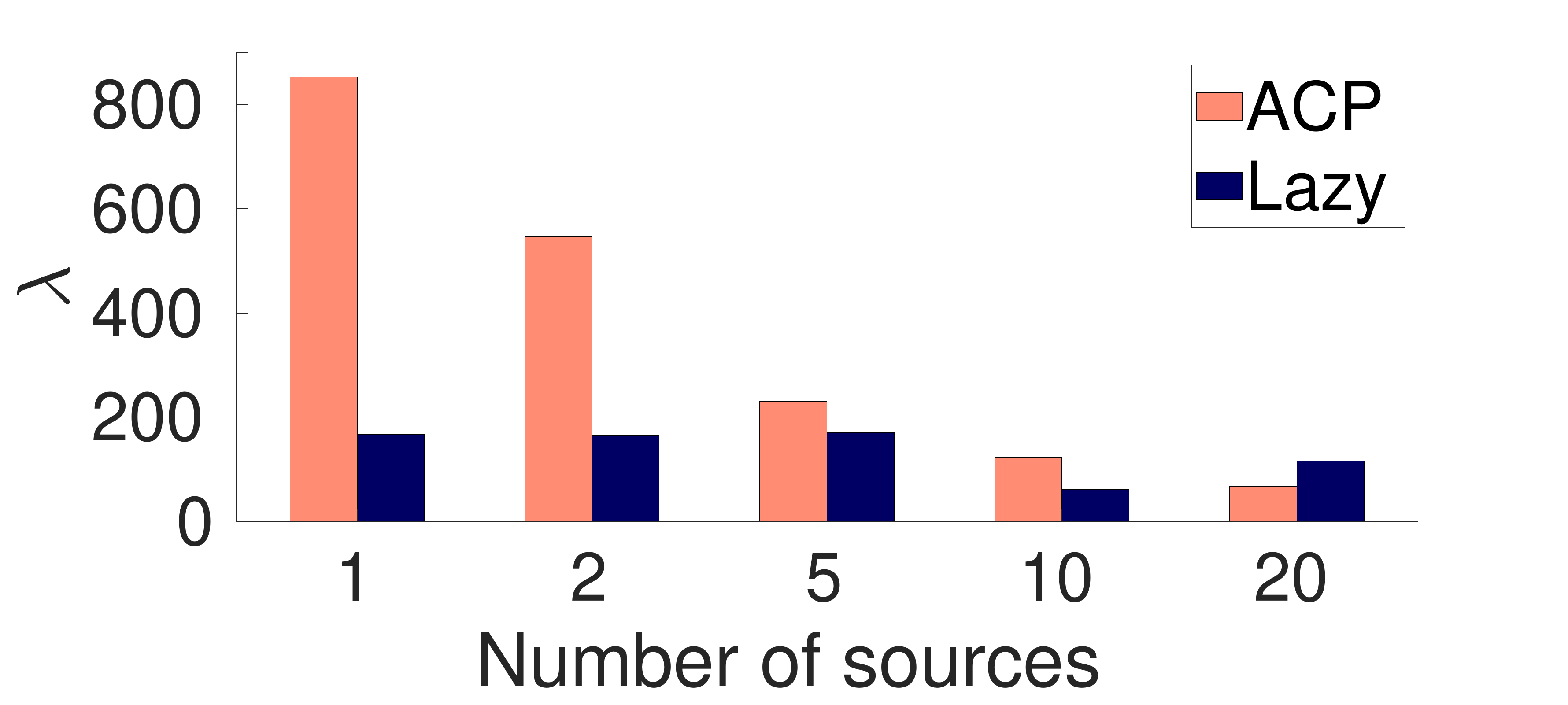}
			\label{fig:lambda_topology_1_fast}}
		\caption{(a) Average source age (b) Average source backlog (c) Average source RTT and (d) Update rate $\lambda$ for Lazy and ACP when all links other than wireless access are $6$ Mbps. All sources used a WiFi PHY rate of $54$ Mbps. The sources are spread over an area of $100$ m$^2$. The standard deviation of shadowing was set to $4$ dB.}
		\label{fig:ageAndBacklogFastAndSlowNets}
	\end{center}
\vspace{-0.2in}
\end{figure}




Figure~\ref{fig:ageAndBacklogFastAndSlowNets} compares the average age, source update rate $\lambda$, the RTT, and the average backlog, obtained when using ACP and \emph{Lazy}. We vary the number of sources in the network from $1$ to $20$. For smaller numbers of sources, the backlog (see Figure~\ref{fig:backlog_topology_1_fast}) per source maintained by ACP is high. This is because, given the similar rate P2P links and higher rate WiFi link, when using ACP, the sources attempt to have their update packets in the queues of the access points and routers in the network. On the other hand, a source using \emph{Lazy} sticks to sending just one packet every RTT on an average. Thus, the average backlog per source stays similar for different numbers of sources. 

As the numbers of sources become large in comparison to the number of hops (six) in the network, even at an average backlog of about $1$ update per source, there is little value in a source sending more than one update per RTT. Note that there are only $6$ hops (queues) in the network. When there are five or more sources, a source sending at a rate faster than $1$ every RTT will have its updates waiting for each other to finish service. This results in ACP maintaining a backlog close to \emph{Lazy} when the numbers of sources are $5$ and more. 

Figure~\ref{fig:lambda_topology_1_fast} shows the average source rate of sending update packets. Observe that the average source rate drops in proportion to the number of sources. While the source rate is about $800$ updates/second when there is only a single source, it is about $70$ when the wireless access is shared by $20$ sources. This scaling down is further evidence of ACP adapting to the introduction of larger numbers of sources. While a source using ACP ramps down its update rate from $800$ to $70$, \emph{Lazy} more or less sticks to the same update rate throughout.

\begin{figure}[!t]             
	\begin{center}
		\subfloat[]{\includegraphics[width=0.25\textwidth]{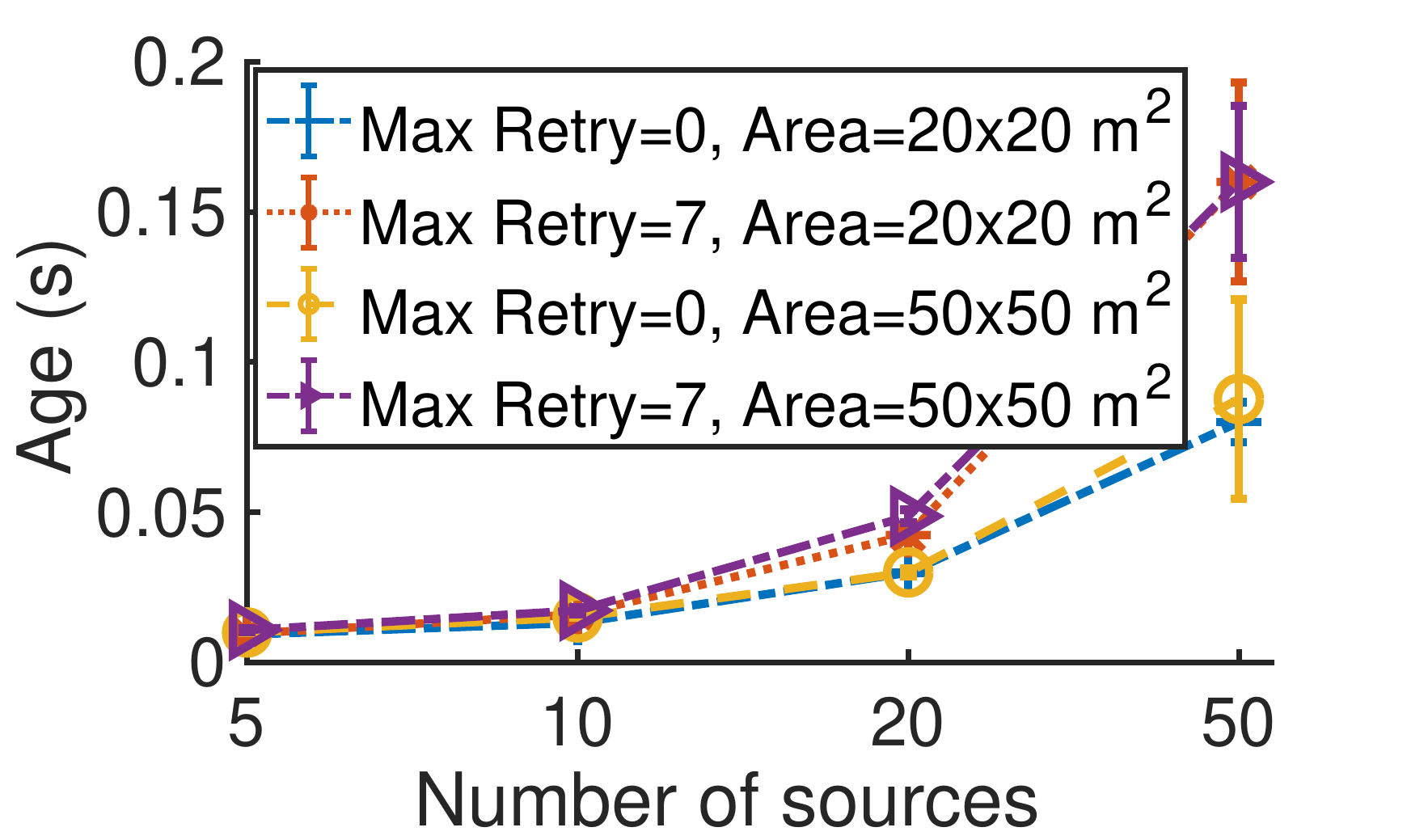}		\label{fig:acpOverWiFi_age400}}
		\hspace{-.25in}
		\subfloat[]{\includegraphics[width=0.25\textwidth]{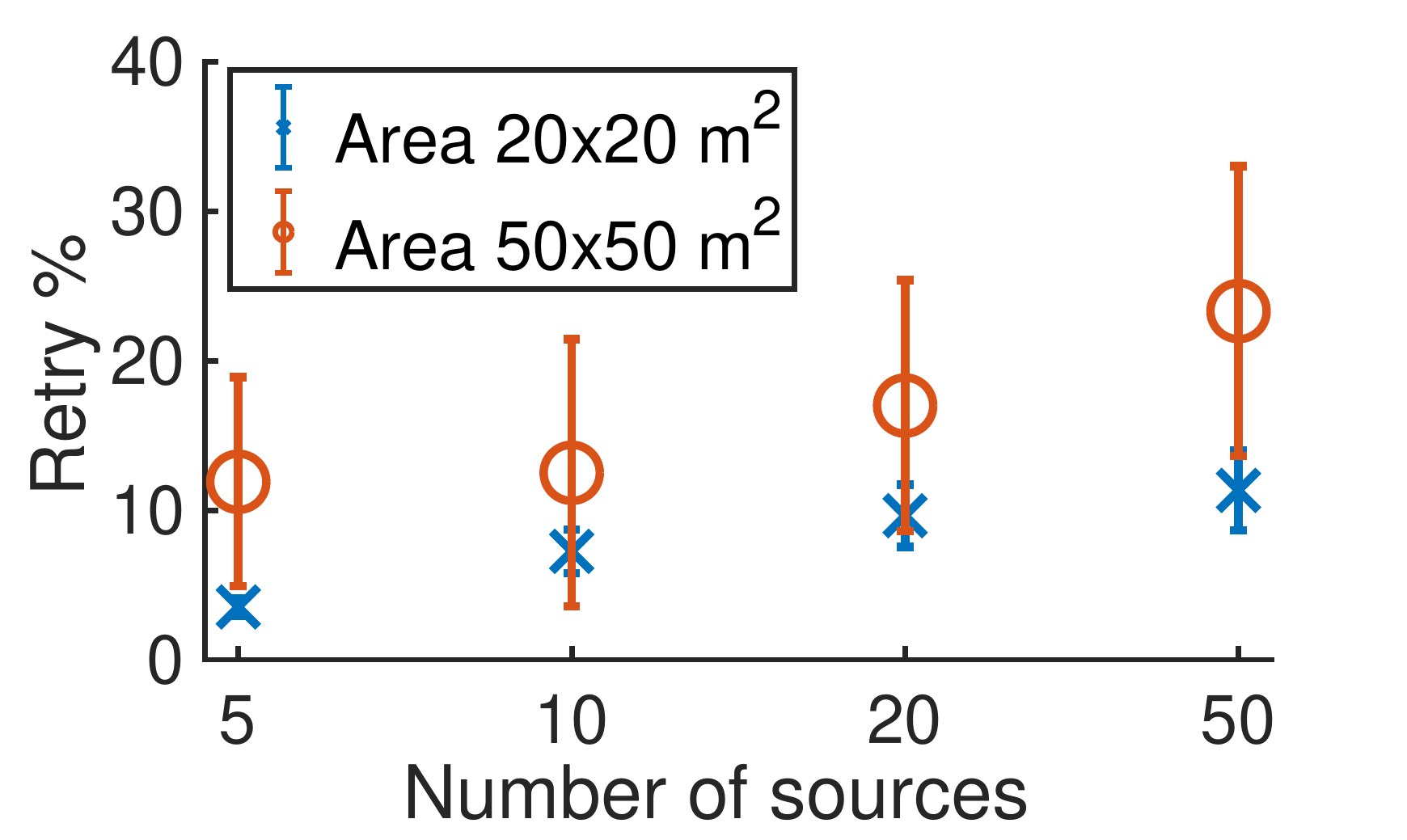}		\label{fig:acpOverWiFi_age_retry}}
		\caption{(a) Age and (b) retry rate as a function of number and density of sources and maximum retry limit. The vertical bars denote a region of $\pm 1$ standard deviation around the mean (marked).}
		\label{fig:acpOverWiFi_age}
	\end{center}
	\vspace{-0.1in}
\end{figure}

\begin{figure}[!t]             
	\begin{center}
		\includegraphics[width=0.48\textwidth]{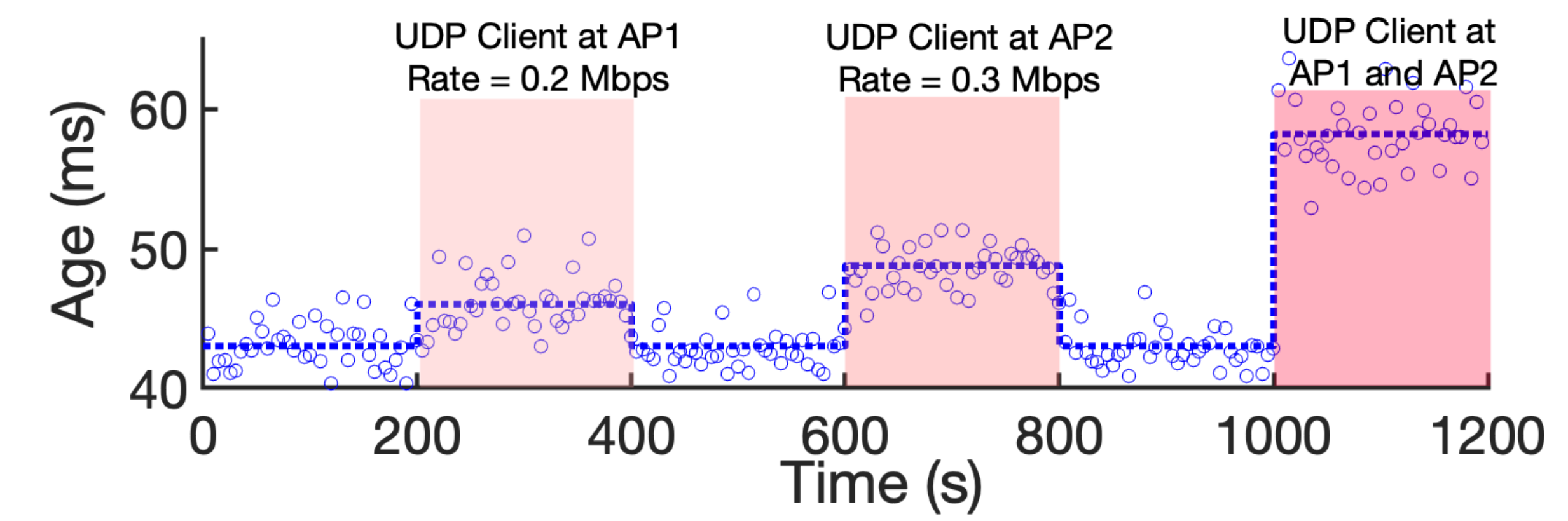}
		\caption{ACP adapts to network changes. Blue circles show the achieved age by an ACP client over time. A UDP client of rate $0.2$ Mbps is connected to AP-1 at $200-400$ secs and $1000-1200$ secs. Another UDP client of rate $0.3$ Mbps is connected to AP-2 at $600-800$ secs and $1000-1200$ secs. A darker shade of pink signifies a larger aggregate UDP load on the network.}
		\label{fig:acpAdapts}
	\end{center}
	\vspace{-0.2in}
\end{figure}

\begin{figure*}[!t]             
	\begin{center}
		\subfloat[]{\includegraphics[width=.33\textwidth]{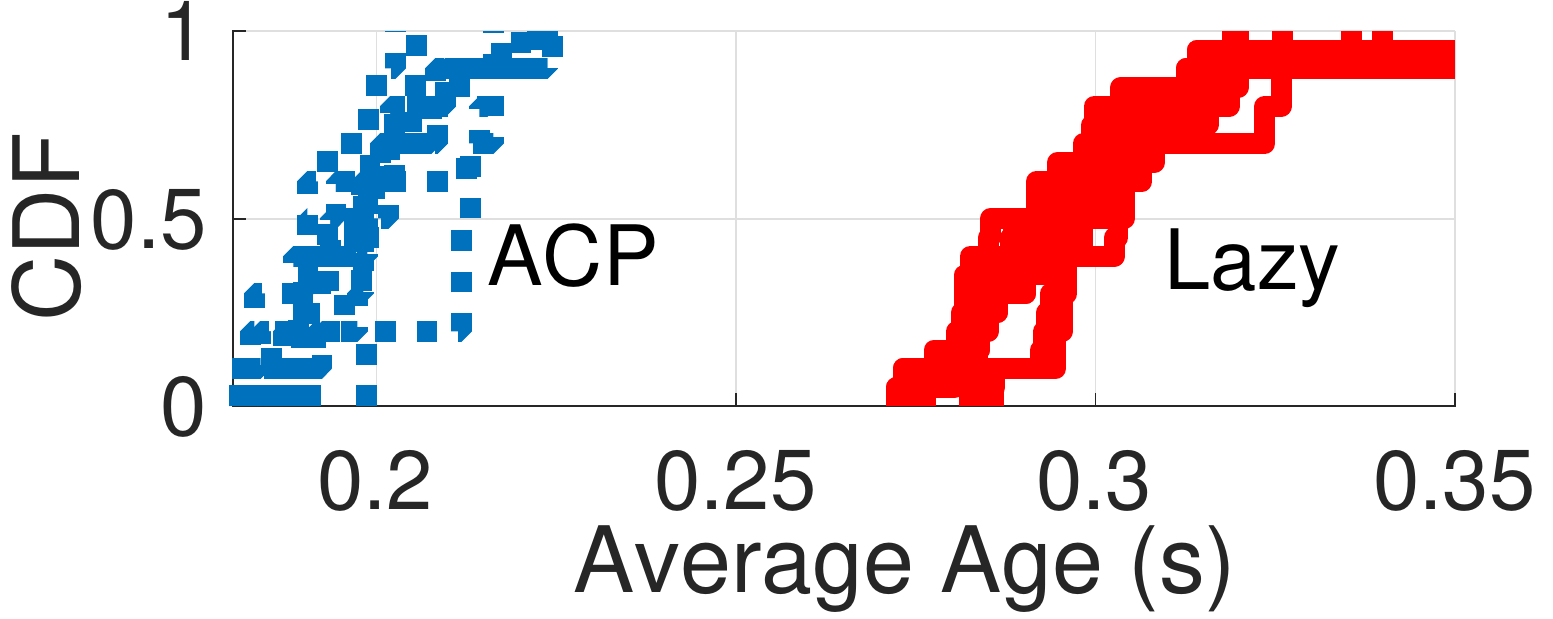}
			\label{fig:cdfAgeRealWorld}}
		\subfloat[]{\includegraphics[width=.33\textwidth]{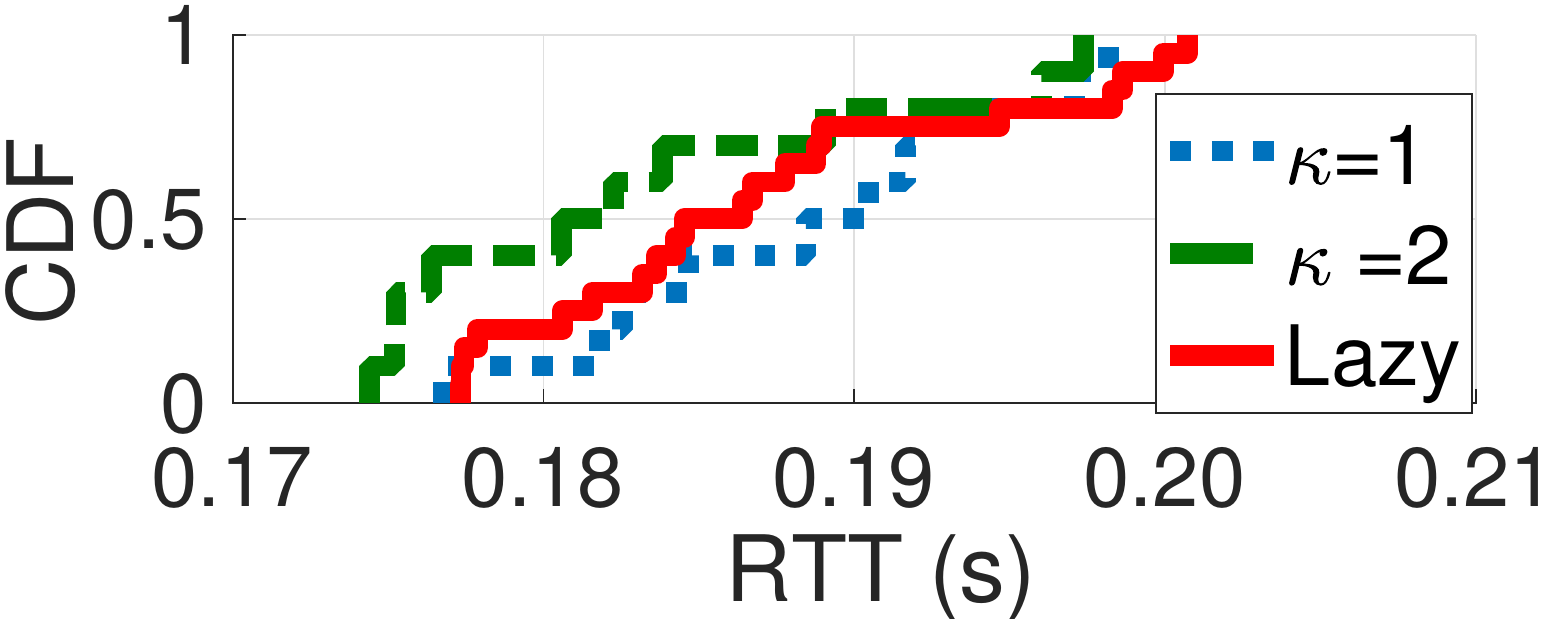}
			\label{fig:cdfRTTRealWorld}}
		\subfloat[]{\includegraphics[width=.33\textwidth]{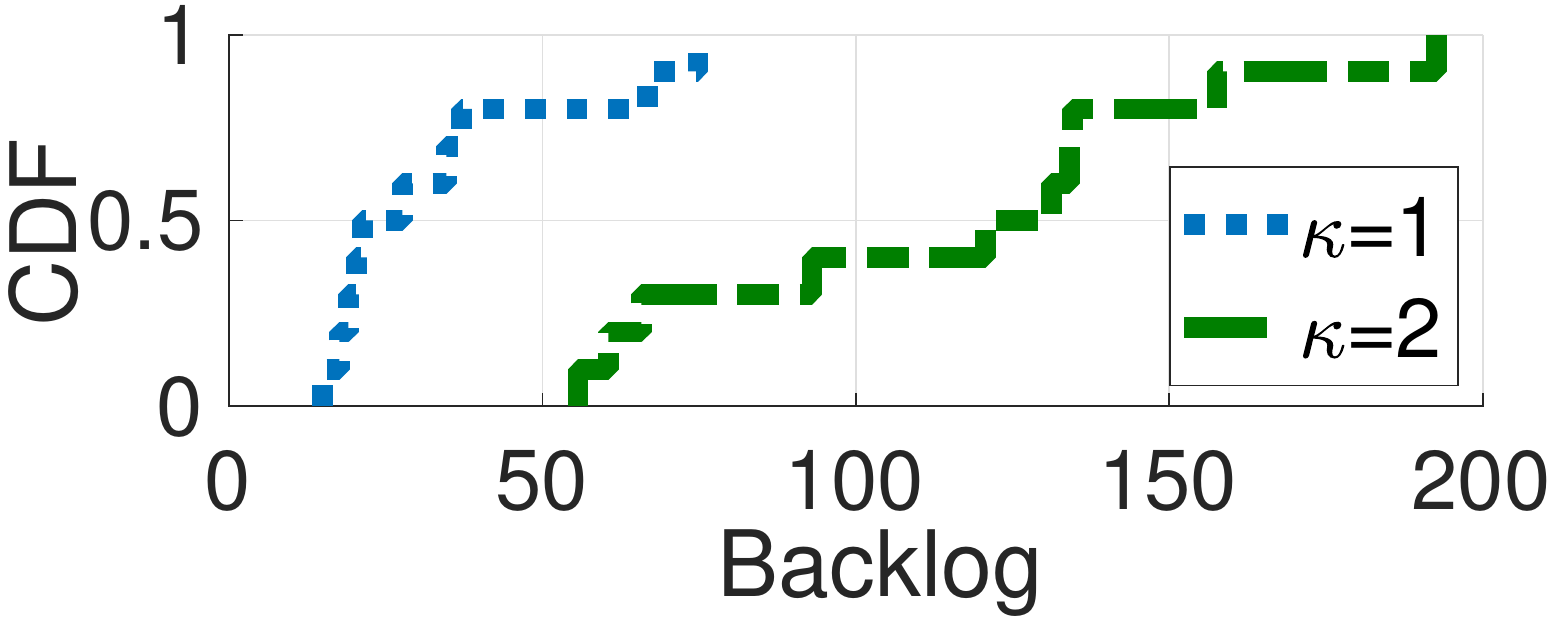}
			\label{fig:cdfBacklogRealWorld}}
		\caption{We compare the CDF(s) of average (a) Age (b) RTT and (c) Backlog obtained over $10$ runs each of \emph{Lazy} and ACP with step size choices of $\kappa=1,2$. The Age CDF(s) of all the $10$ sources are shown.}
		\label{fig:CDFFromRealWorld}
	\end{center}
\vspace{-0.14in}
\end{figure*}
\begin{figure}[!t]             
	\begin{center}
		\subfloat[]{\includegraphics[width=.24\textwidth]{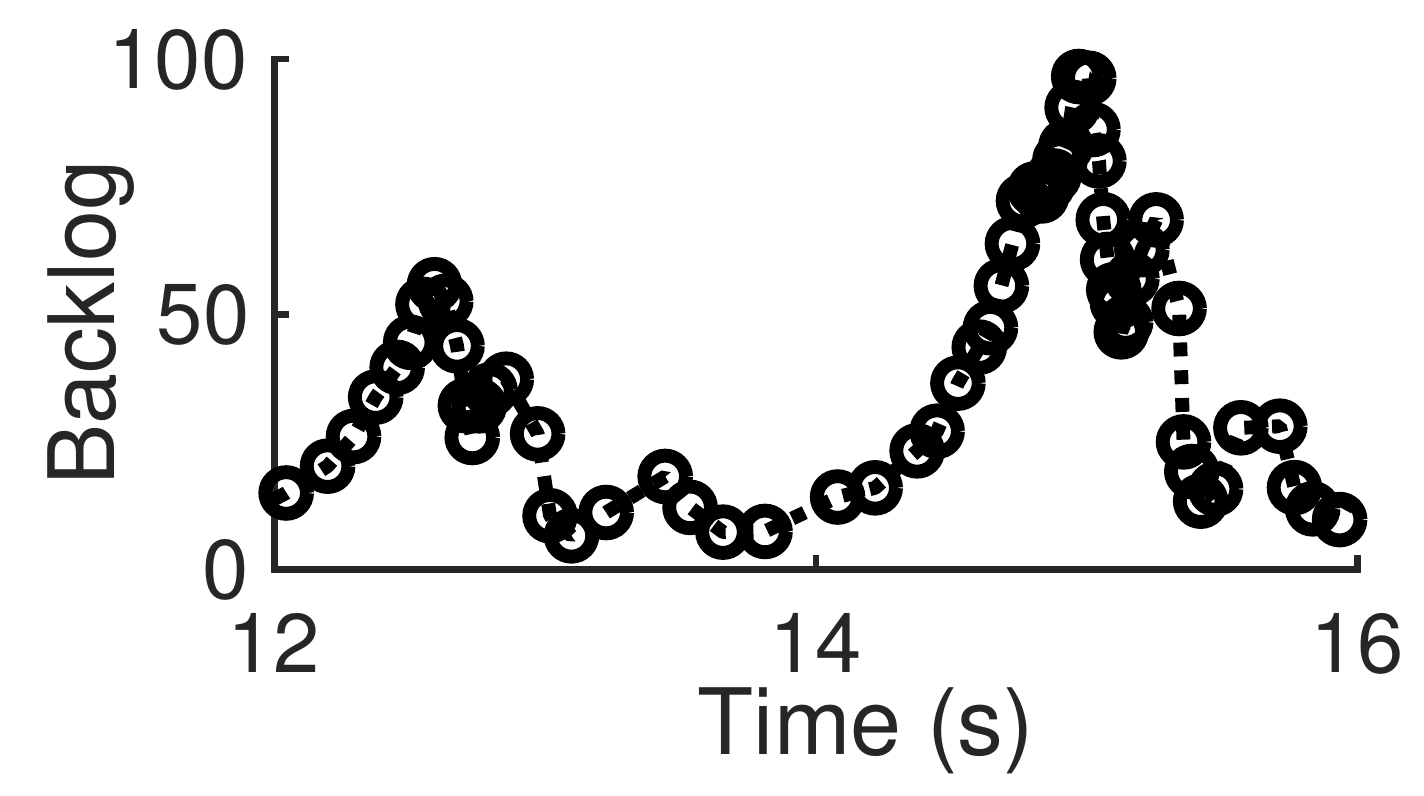}
			\label{fig:acpBacklogEvol}}
		\subfloat[]{\includegraphics[width=.24\textwidth]{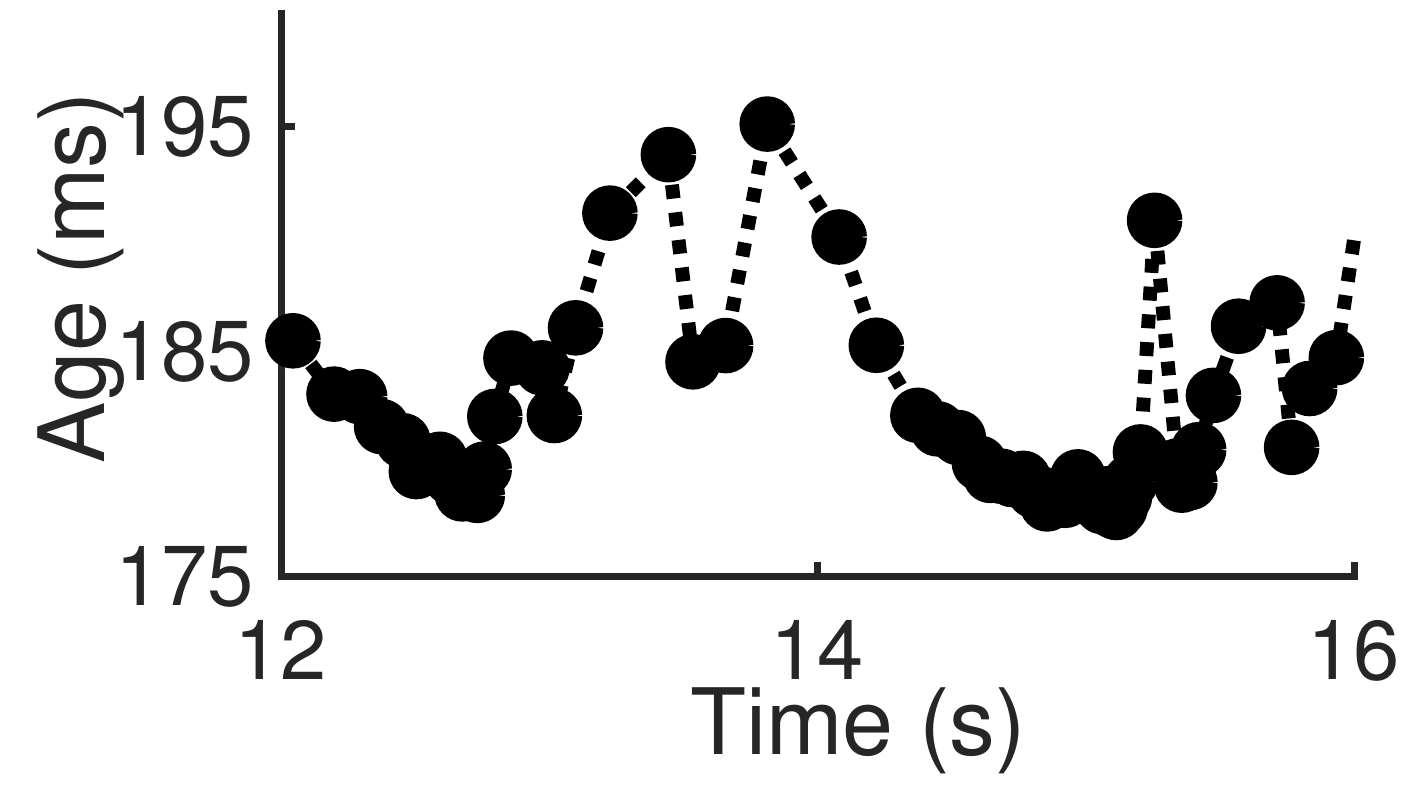}
			\label{fig:acpAgeEvol}}
		\caption{The time evolution of average backlog and age that resulted from one of the ACP source sending updates over the Internet.}
		\label{fig:acpEvolution}
	\end{center}
\vspace{-0.2in}
\end{figure}

This artificial constraint of a very few hops combined with a single end-to-end path is removed in the real-world experiments that we present in the next section. As we will see, sources accessing a common access point will maintain high backlogs over their end-to-end connections to their monitors.

The absolute improvements in average age achieved by ACP, see Figure~\ref{fig:age_topology_1_fast}, for fewer numbers of sources seem nominal but must be seen in light of the fact that end-to-end RTT of the simulated network under light load conditions is very small (about $5$ msec as seen in Figure~\ref{fig:rtt_topology_1_fast}). ACP achieves a $21\%$ and $13\%$ reduction in age with respect to Lazy, respectively, for a single source and two sources.

The only impact that changing the link rates of the P2P links had was a corresponding change in RTT and Age. For example, while the average age achieved by a source using ACP in a $20$ source network with P2P link rates $0.3$ Mbps was $\approx 6$ seconds, it was $\approx 0.25$ seconds when the P2P link rates were set to $6.0$ Mbps. The larger RTT for the latter meant smaller $\lambda$ of about $5$ updates/second/source. The backlogs, as one would expect, were similar, however.

Next consider Figure~\ref{fig:acpOverWiFi_age400} that shows the impact of maximum allowed retries, numbers of sources (varied from $5$ to $50$), and source density (areas of $50\times 50$ m$^2$ and $20\times 20$ m$^2$), on average age. The standard deviation of shadowing was set to $12$. Note that age is similar for the two simulated areas for a given setting of maximum retries. However, it is significantly larger for when the max retry limit is set to $7$ in comparison to when no retries are allowed. This is especially true when the network has larger numbers of sources. Larger numbers of sources witness higher rates of retries (Figure~\ref{fig:acpOverWiFi_age_retry}, retry limit is $7$) due to a higher rate of packet decoding errors that result from collisions over the WiFi medium access shared by all sources. Retries create a two-fold problem. One that a retry may keep a fresher update from being transmitted. Second, ACP, like TCP, confuses packet drops due to channel errors to be network congestion. This causes it to unnecessarily reduce $\lambda$ in response to packet errors, which increases age. In summary, retries at the wireless access are detrimental to keeping age low. Finally, observe in Figure~\ref{fig:acpOverWiFi_age400} that the spread of ages achieved by sources is very small. In fact, we see that sources in a network achieve similar ages and in all our simulations the Jain's fairness index~\cite{jain1999throughput} was found to be close to the maximum of $1$. 


ACP adapts rather quickly to the introduction of other flows that congest the network. This is exemplified by Figure~\ref{fig:acpAdapts}. We introduced one to two UDP flows at different points in the network used for simulation (Figure~\ref{fig:simulationNetwork}), where all links are $1$ Mbps. ACP reduces $\lambda$ appropriately and adapts backlog to desired levels.

\section{Inter-Continental Updates}
\label{sec:realWorldResults}
We will show results for when $10$ sources sent their updates to monitors on a server in another continent. The sources, as described earlier, gained access to the Internet via an enterprise access point. The results were obtained by running ACP and \emph{Lazy} alternately for $10$ runs. Each run was restricted to $1000$ update packets long so that on an average ACP and \emph{Lazy} experienced similar network conditions. We ran ACP for $\kappa=1$ and $\kappa=2$. Using traceroute, we observed that the number of hops was large, about $30$, during these experiments.

Figure~\ref{fig:CDFFromRealWorld} summarizes the comparison of ACP and \emph{Lazy}. Figure~\ref{fig:cdfAgeRealWorld} shows the cumulative distribution functions (CDF) of the average age obtained by each source when using ACP (using $\kappa = 1$) and the corresponding CDF(s) when using \emph{Lazy}. As is seen in the figure, ACP outperforms \emph{Lazy} and obtains a median improvement of about $100$ msec in age ($\approx 33\%$ over average age obtained using \emph{Lazy}). This over an end-to-end connection with median RTT of about $185$ msec. Further, observe that the age CDF(s) for all the sources when using either ACP or \emph{Lazy} are similar. This hints at sources sharing the end-to-end connection in a fair manner. Also, observe from Figure~\ref{fig:cdfRTTRealWorld} that the median RTT(s) for both ACP and \emph{Lazy} are almost the same. This signifies that ACP maintains a backlog of update packets in a manner such that the packets don't suffer additional delays because multiple packets of the source are traversing the network at the same time.

Lastly, consider a comparison of the CDF of average backlogs shown in Figure~\ref{fig:cdfBacklogRealWorld}. ACP exploits very well the fast end-to-end connection with multiple hops and achieves a very high median average backlog of about $30$ when using a step size of $1$ and a much higher backlog when using a step size of $2$. We observe that step size $\kappa=1$ worked best age wise. \emph{Lazy}, however, achieves a backlog of about $1$ (not shown).

We end by showing snippets of ACP in action over the end-to-end path. Figures~\ref{fig:acpBacklogEvol} and~\ref{fig:acpAgeEvol} show the time evolution of average backlog and average age, as calculated at control epochs. ACP increases backlog in small steps (see Figure~\ref{fig:acpBacklogEvol}, $14$ seconds onward) over a large range followed by a rapid decrease in backlog. The increase coincides with a reduction in average age, and the rapid decrease is initiated once age increases. Also, observe that age decreases very slowly (dense regions of points low on the age curve around the $15$ second mark) with an increase in backlog just before it increases rapidly. The region of slow decrease is around where, ideally, backlog must be set to keep age to a minimum.

\section{Conclusions}
\label{sec:conclusions}
We proposed the Age Control Protocol, which is a novel transport layer protocol for real-time monitoring applications that desire the freshness of information communicated over the Internet. ACP works in an application-independent manner. It regulates the update rate of a source in a network-transparent manner. We detailed ACP's control algorithm that adapts the rate so that the age of the updates at the monitor is minimized. Via network simulations and real-world experiments, we showed that ACP adapts the source update rate well to make an effective use of network resources available to the end-to-end connection between a source and its monitor.

\balance
\section*{Acknowledgment}

This research was funded by TCS Research Scholarship Program and Young Faculty Research Fellowship (Visvesvaraya Ph.D. scheme) from MeitY, Govt. of India.


\bibliographystyle{abbrv} 
\bibliography{IEEEtran,AOI-test}

\end{document}